\documentclass[twocolumn,10pt,aps,prd]{revtex4-2}
\usepackage{hyperref,graphicx,amsfonts,amsmath,amssymb}
\usepackage[latin1]{inputenc}
\usepackage{tikz}
\usetikzlibrary{trees}
\usetikzlibrary{decorations.pathmorphing}
\usetikzlibrary{decorations.markings}
\tikzset{
    photon/.style={decorate, decoration={snake}, draw=red, solid},
    electron/.style={draw=blue, solid, postaction={decorate},
        decoration={markings,mark=at position .55 with {\arrow[draw=blue]{<}}}},
    positron/.style={draw=blue, solid, postaction={decorate},
        decoration={markings,mark=at position .55 with {\arrow[draw=blue]{>}}}},
    Higgs/.style={draw=green, dashed, postaction={decorate},
        decoration={markings,mark=at position .55 with {\arrow[draw=green]{>}}}},
    gluon/.style={decorate, draw=magenta,
        decoration={coil,amplitude=4pt, segment length=5pt}}
}

\begin{document}

\title{Parameter dependence and analysis of the 2HDM neutral Higgs boson pair production and decay at future lepton colliders}

\author{Majid Hashemi}
\email{majid.hashemi@cern.ch}
\author{Neda Nowbakht Ghalati}
\email{neda.nobakht@yahoo.com}
\affiliation{Physics Department, College of Sciences, Shiraz University, Shiraz, 71946-84795, Iran}
%...........................................................
\begin{abstract}
In this work, we present a study of the neutral Higgs bosons in the two Higgs doublet model (2HDM) in terms of their production processes and decay channels as a function of the model parameters. The analysis is performed for all four types of the 2HDM and the most promising processes and decay channels are identified for each type of the model. Several Higgs boson mass scenarios below and above the threshold of decay to gauge boson pair are introduced and the corresponding categories of final states are analyzed. The event analysis including collider beam spectrum and detector simulation shows that future lepton colliders have the potential to explore regions in the 2HDM parameter space, which have not yet been excluded by LHC, in a few weeks of operation. Final results are presented in terms of the signal distributions on top of the background and 95$\%$ CL exclusion and 5$\sigma$ contours based on center of mass energy of $\sqrt{s}=500$ GeV at the Compact Linear Collider (CLIC) or International Linear Collider (ILC).
\end{abstract}
\maketitle
%...................................................................
\section*{Introduction}
One of the main achievements in high energy physics in the last decade is the observation of a new boson at the Large Hadron Collider (LHC) by the two collaborations ATLAS and CMS \cite{HiggsATLAS,HiggsCMS}.

The observed particle is the candidate for the missing key element of the standard model, i.e., the Higgs boson, $\mathit{h_{SM}}$ \cite{Higgs1,Higgs2,Higgs3,Englert1,Kibble1,Kibble2} and its properties are in reasonable agreement with SM predictions as verified by various analyses at the LHC \cite{LHC1,LHC2,LHC3,LHC4,LHC5,LHC6,LHC7,LHC8}.

Within the uncertainty of these measurements, there is still possibility to consider beyond Standard Model (BSM) such as the two Higgs doublet model (2HDM) \cite{2hdm1,2hdm2,2hdm3} which introduces the SM-like Higgs boson candidate together with extra neutral and charged Higgs bosons.

Although the 2HDM provides the Higgs sector for supersymmetry in the minimal form (MSSM) \cite{MSSM1,MSSM2,MSSM3}, it is still attractive as a standalone model due to the possibility of better agreement with experimental data \cite{fmahmoudi}.

The structure of the 2HDM and its parameters provide the possibility to coincide the lightest Higgs boson properties to those of the SM Higgs boson \cite{2hdm_theorypheno}. The heavy neutral CP-even(CP-odd) Higgs bosons $\mathit{H}$($\mathit{A}$) and the two charged Higgs bosons $\mathit{H^{\pm}}$ are considered as extra Higgs bosons to be observed or excluded in the current or future experiments.

After the discovery of the light Higgs boson candidate, one of the main goals of the ATLAS and CMS collaborations has been the search for the extra Higgs bosons.

The ATLAS collaboration has reported an analysis of $pp$ $\rightarrow A\rightarrow Zh$ \cite{atlas3} where the CP-odd Higgs boson, $A$, decays to $Z$ boson and 125 GeV Higgs boson. They cover four types of 2HDM based on the Higgs-fermion couplings and results are presented in terms of exclusion contours in the parameter space. These results are confirmed by the CMS collaboration \cite{cms2}. The heavy Higgs conversion, i.e., $A \to ZH$ has been analyzed by the two collaborations CMS \cite{cms1} and ATLAS \cite{atlas1,atlas2}. We will discuss about these results in the next sections.

While collision data is taken by the two LHC collaborations CMS and ATLAS, there are ongoing analyses focusing on the possibility of observing extra Higgs bosons at the LHC luminosity upgrade \cite{hllhc1,hllhc2} and also future lepton colliders such as CLIC \cite{clichiggs1,clichiggs2}, ILC \cite{ilchiggs}, FCC \cite{fcchiggs} and CEPC \cite{cepchiggs}.

In a number of recent works, we analyzed charged \cite{hashemimuon1,hashemimuon2,hashemichlc} and neutral \cite{htype1,htype3,htype4} Higgs boson production and decay at lepton colliders and provided prospects for their observation in different scenarios using benchmark points in the parameter space.
The above analyses were based on the alignment limit \cite{align1,align2,align3} which is defined as the scenario in which the properties of one of the neutral CP-even Higgs mass eigenstates coincide with those of the SM Higgs boson.

The alignment limit is naturally achieved in the so-called decoupling limit where the masses of other scalar states are large and decouple from the SM-like Higgs boson \cite{decoupling}. However, it is possible to achieve the alignment limit even without decoupling \cite{align1,align2,align3} which has been the case in our previous studies.

In the current work, we consider the possibility of migrating from the alignment limit and we perform a general scan of the parameter space to analyze the neutral Higgs boson branching ratio of decays. The analysis is not limited to a specific type of the 2HDM and all types are analyzed and compared to reach a conclusion on the choice of the most relevant production process and decay channel for the neutral Higgs bosons in each part of the parameter space.

In what follows, a brief theoretical description of the 2HDM and the software setup used for the analysis are presented. Next, we discuss about the signal processes adopted by LHC collaborations and then present our detailed study of the neutral Higgs boson decay channels in different mass scenarios, theoretical constraints and their relevance in each type of the model. The final conclusion for each type of the 2HDM is presented based on kinematic analysis of events.

\section{The two Higgs doublet model}
The SM Higgs Lagrangian is written in the form
\begin{equation}
 \mathcal{L}=(\partial_{\mu}\Phi)^\dagger (\partial^{\mu}\Phi)-\mathcal{V}(\Phi)
\end{equation}
where $\mathcal{V}$ is the Higgs potential based on only one Higgs doublet $\Phi$:
\begin{equation}
\mathcal{V} = \mu^2\Phi^\dagger\Phi+\lambda(\Phi^\dagger\Phi)^2
\end{equation}

With this form of the potential, the condition to have non-zero vacuum expectation value for the Higgs field is $\mu^2<0$.

The two Higgs doublet model is made as an extension of the SM Higgs sector by introducing two Higgs doublets $\Phi_1$ and $\Phi_2$.

Writing all possible Lagrangian terms requires additional degrees of freedom. The SM Higgs potential $\mu^2$ term is extended to include three parameters $m_{11}^2$, $m_{22}^2$ and $m_{12}^2$ and the $\lambda$ term is extended to seven terms containing $\lambda_1$ to $\lambda_7$ \cite{decoupling,2hdm_higgssector1,tanbsignificance}. In such a general Higgs potential, Higgs-boson-associated FCNC interactions exist. It has been shown that such FCNC terms are avoided at tree level by imposing discrete $Z_2$ symmetry ($\Phi_1 \to \Phi_1$ and $\Phi_2 \to -\Phi_2$) \cite{2hdm2}. The 2HDM Higgs potential under softly broken $Z_2$ symmetry (allowing $m_{12}\neq 0$) reduces to the following form \cite{2hdm_higgssector2}:
\begin{align}
\mathcal{V} \nonumber &= m_{11}^2\Phi_1^\dagger\Phi_1+m_{22}^2\Phi_2^\dagger\Phi_2-m_{12}^2\left(\Phi_1^\dagger\Phi_2+\Phi_2^\dagger\Phi_1\right)\\
    \nonumber &+\frac{1}{2}\lambda_1\left(\Phi_1^\dagger\Phi_1\right)^2+\frac{1}{2}\lambda_2\left(\Phi_2^\dagger\Phi_2\right)^2\\ \nonumber
    \nonumber &+\lambda_3\left(\Phi_1^\dagger\Phi_1\right)\left(\Phi_2^\dagger\Phi_2\right)+\lambda_4\left(\Phi_1^\dagger\Phi_2\right)\left(\Phi_2^\dagger\Phi_1\right)\\ \nonumber
    \nonumber &+\frac{1}{2}\lambda_5\left[\left(\Phi_1^\dagger\Phi_2\right)^2+\left(\Phi_2^\dagger\Phi_1\right)^2\right]\\ \nonumber
  \label{lag}
\end{align}
The condition corresponding to the SM $\mu^2<0$ is that the Higgs mass matrix made of $m^2_{ij}$ has at least one negative eigenvalue. If this is the case, the two doublets can be written in terms of their vacuum expectation values:
\begin{equation}
\langle \Phi_1 \rangle=\frac{1}{\sqrt{2}} \left(
\begin{array}{c} 0\\ v_1\end{array}\right), \qquad \langle
\Phi_2\rangle=
\frac{1}{\sqrt{2}} \left(\begin{array}{c}0\\ v_2
\end{array}\right)\,\label{vevs}
\end{equation}
where the ratio of the two vevs is a free parameter of the model denoted by $\tan\beta=v_2/v_1$ with $v^2=v_1^2+v_2^2=(246~ \textnormal{GeV})^2$.

The other parameter is the mixing angle $\alpha$ used to diagonalize the CP-even Higgs mass-squared matrix. The two parameters $\alpha$ and $\beta$ appear in the Higgs-fermion and Higgs-gauge couplings \cite{2hdm_higgssector2}.
The Higgs-fermion Yukawa interactions, keeping only the neutral Higgs interactions, take the following form
\begin{equation}
\mathcal{L}_{Y}=\sum_{f=u,d,\ell}\frac{m_f}{v}\Big(\xi_{h}^{f}\bar{f}fh + \xi_{H}^{f}\bar{f}fH - i\xi_{A}^{f}\bar{f}\gamma_5fA \Big)
\label{yukawa1}
\end{equation}
where the couplings are expressed in terms of the corresponding SM value, $m_f/v$, times the type dependent factors $\xi_{h/H/A}^{u,d,\ell}$ presented in Tab. \ref{couplings}. The CP-even Higgs coupling terms are sometimes written in terms of $\sin(\beta-\alpha)$ or $\cos(\beta-\alpha)$ using trigonometric relations \cite{decoupling}:
\begin{align}
\sin\alpha/\sin\beta \nonumber&~=~ \cos(\beta-\alpha)-\cot\beta\sin(\beta-\alpha)\\ \nonumber
\cos\alpha/\cos\beta &~=~ \cos(\beta-\alpha) + \tan\beta\sin(\beta-\alpha)\\\nonumber
-\sin\alpha/\cos\beta &~=~ \sin(\beta-\alpha)-\tan\beta\cos(\beta-\alpha)\\\nonumber
\cos\alpha/\sin\beta &~=~ \sin(\beta-\alpha) + \cot\beta\cos(\beta-\alpha)\\
\label{trig}
\end{align}
The Higgs boson couplings to gauge bosons are model independent and, normalized to their corresponding SM values, are
\begin{equation}
 g_{hVV}=\sin(\beta-\alpha),~~g_{HVV}=\cos(\beta-\alpha).
\label{hgauge}
 \end{equation}
There is no tree level coupling of the CP-odd Higgs boson $A$ to vector bosons.

Since $\xi^{u,d,\ell}_h$ is either $\cos\alpha/\sin\beta$ or $-\sin\alpha/\cos\beta$, it is obvious through Eq. \ref{trig} and \ref{hgauge} that both $h$-fermion and $h$-gauge couplings align to their corresponding SM values if $\sin(\beta-\alpha)=1$. One of the consequences of the alignment is that the heavier CP-even Higgs coupling to gauge bosons vanishes and its couplings to fermions is expressed in terms of $\tan\beta$ or $\cot\beta$.

The above simplified scheme of Higgs-fermion/gauge couplings has been analyzed in various analyses. The extra Higgs bosons ($H$ and $A$) are gaugeophobic and their couplings to fermions, normalized to the corresponding SM couplings, depend on $\beta$ while $\alpha$ is fixed through $\sin(\beta-\alpha)=1$.

In this work, we do not restrict ourselves to the above requirement and instead, we take $\sin(\beta-\alpha)$ and $\tan\beta$ as input to evaluate the couplings of Tab. \ref{couplings} with the use of \texttt{2HDMC 1.8} \cite{2hdmc1,2hdmc2,2hdmc3}. The full combination of experimental limits are also obtained from the LHC 13 TeV run analyses using \texttt{HiggsBounds 5.10.2} \cite{hb1,hb2,hb3,hb4,hb5} and \texttt{Higgs Signals 2.6.2} \cite{hs1,hs2,hs3} to make sure that heavy neutral Higgs boson masses and the parameters used for the event analysis are allowed. The SM Higgs boson measurements constraints are shown in all scenarios based on the results reported in \cite{atlas:2020}. In addition to the experimental limits, the theoretical constraints of potential stability (positivity) \cite{pos1,pos2,pos3,pos4,pos5}, unitarity and perturbativity \cite{uni1,uni2,uni3} are also verified.
\begin{table}[h]
\begin{center}
\begin{tabular}{|c|c|c|c|c|}  \hline
{} & Type 1  & Type 2 & Type 3 & Type 4 \\
\hline
$\xi^u_h $ & $\cos\alpha/\sin\beta$ & $\cos\alpha/\sin\beta$ & $\cos\alpha/\sin\beta$ & $\cos\alpha/\sin\beta$ \\
\hline
$\xi^d_h $ & $\cos\alpha/\sin\beta$ & $-\sin\alpha/\cos\beta$ & $-\sin\alpha/\cos\beta$  & $\cos\alpha/\sin\beta$ \\
\hline
$\xi^\ell_h $ & $\cos\alpha/\sin\beta$ & $-\sin\alpha/\cos\beta$ & $\cos\alpha/\sin\beta$   & $-\sin\alpha/\cos\beta$   \\
\hline\
$\xi^u_H $ & $\sin\alpha/\sin\beta$ & $\sin\alpha/\sin\beta$   & $\sin\alpha/\sin\beta$ &$\sin\alpha/\sin\beta$    \\
\hline
$\xi^d_H $ & $\sin\alpha/\sin\beta$ & $\cos\alpha/\cos\beta$   & $\cos\alpha/\cos\beta$ &$\sin\alpha/\sin\beta$    \\
\hline
$\xi^\ell_H $ & $\sin\alpha/\sin\beta$ & $\cos\alpha/\cos\beta$   & $\sin\alpha/\sin\beta$ &$\cos\alpha/\cos\beta$ \\
\hline
$\xi^u_A $ & $\cot\beta$ & $\cot\beta$ & $\cot\beta$ &$\cot\beta$    \\
\hline
$\xi^d_A $ & $- \cot\beta$ & $\tan\beta$ & $\tan\beta$ &$-\cot\beta$  \\
\hline
$\xi^\ell_A $ & $- \cot\beta$ & $\tan\beta$ & $-\cot\beta$  &$\tan\beta$   \\
\hline
\end{tabular}
\end{center}
\caption{Yukawa couplings of the up-type ($u$) and down-type ($d$) quarks and leptons ($\ell$) to the neutral Higgs bosons $h/H/A$ in different types of 2HDM. There are also other names for types 3 and 4: flipped and lepton-specific \cite{2hdm_theorypheno}.}
\label{couplings}
\end{table}

\section{The LHC search channel for 2HDM neutral Higgs bosons}
Before proceeding to our detailed analysis, we discuss about the LHC (ATLAS and CMS) search channel for the 2HDM neutral Higgs bosons presented in \cite{atlas3,cms2,cms1,atlas1,atlas2}.

The above analyses are based on the single CP-odd Higgs boson production followed by the subsequent decay $A \to ZH$ or $A \to Zh_{\mathrm{SM}}$. The CMS collaboration also considers the $m_H>m_A$ possibility through $H \to ZA$ decay \cite{cms1}. In our study, we assume $m_A>m_H$ while the analysis of the opposite case can be performed in a similar way.

The LHC searches for the neutral Higgs bosons are divided into two categories of Higgs boson conversion, i.e., $A\to Zh_{SM}$ and $A \to ZH$ where the final state contains the SM-like Higgs boson or the CP-even heavy Higgs boson.

The decay chains for the two processes are slightly different. The case of $A\to Zh$ involves $\cos(\beta-\alpha)$ as the coupling factor and vanishes at the alignment limit which is defined as $\cos(\beta-\alpha)=0$. While the $A\to Zh$ coupling is type independent, $h_{SM} \to b\bar{b}$ depends on the type of the 2HDM which results in different patterns for the four types of the 2HDM in two dimensional $\tan\beta$ vs $\cos(\beta-\alpha)$ space.

\begin{figure}[h]
\hspace*{-0.1in}
 \includegraphics[height=0.4\textwidth,width=0.5\textwidth]{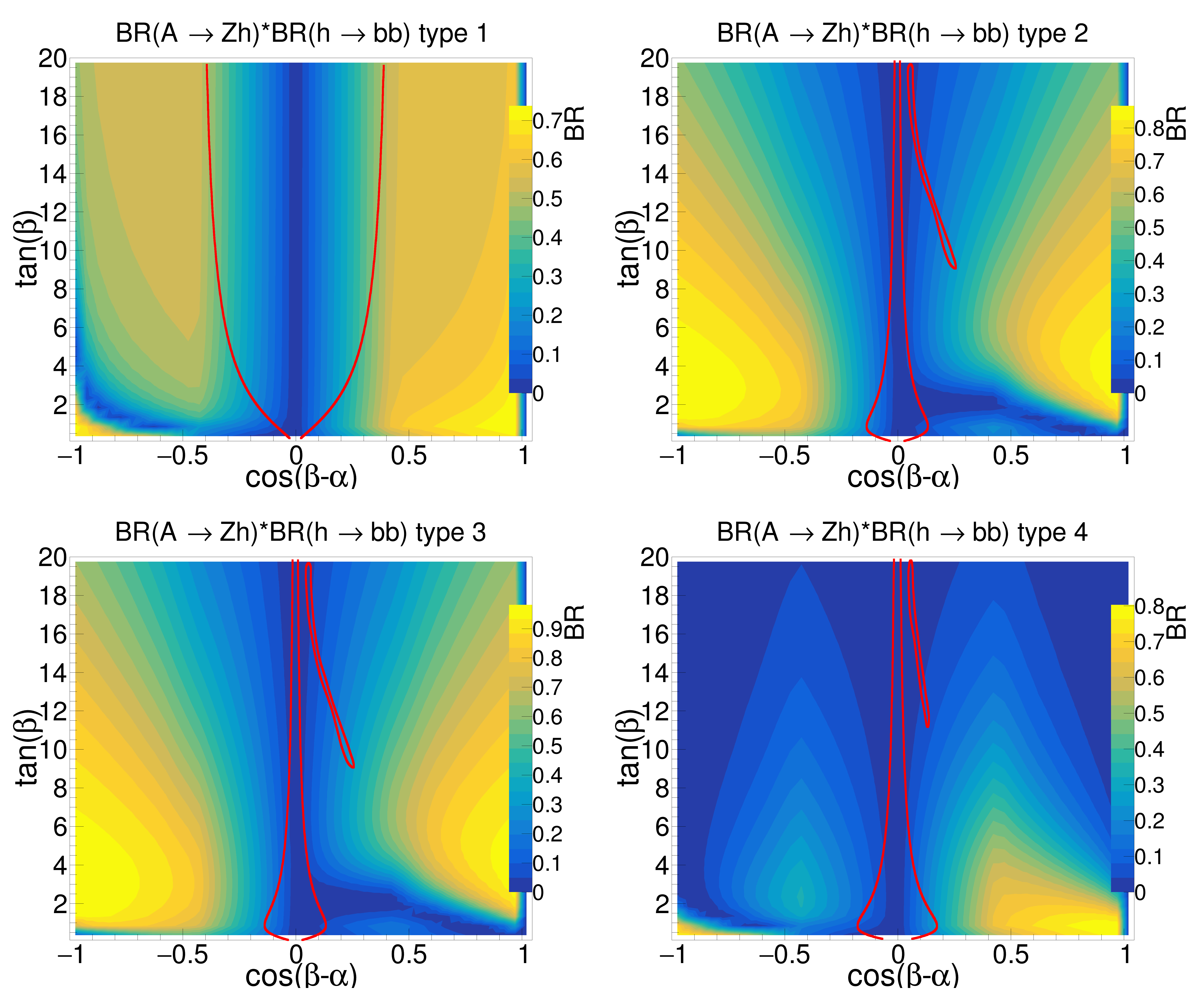}
\caption{The product of the branching ratio of the two decay channels adopted by LHC in \cite{atlas3,cms2}. Here $m_{A/H}~=~300$ GeV. The area outside the red line is excluded by the LHC SM Higgs boson measurements. \label{AZh}}
 \end{figure}

Figure \ref{AZh} shows $\mathrm{BR}(A\to Zh)\times BR(h\to b\bar{b})$ for the four types as a function of $\tan\beta$ and $\cos(\beta-\alpha)$ assuming $m_{H/A}=300$ GeV. Except for the lepton-specific type which is essentially designed for $h\to \ell\ell$, any suppression of the product of branching ratios in $A\to Zh\to Zbb$ which occurs at $\cos(\beta-\alpha)\neq 0$ is due to the suppression of $h\to bb$, otherwise a symmetric pattern around the vertical line of $\cos(\beta-\alpha)=0$ would have been obtained. 

The area inside the red line is allowed and consistent with the LHC light Higgs boson observation. The ATLAS and CMS analyses which used $A\to Zh$ followed by $h\to b\bar{b}$ (\cite{atlas3,cms2}) left the low BR (blue) regions especially the central vertical line of $\cos(\beta-\alpha)=0$ and excluded the rest. 
   
Although Fig. \ref{AZh} is for $m_{H/A}=300$ GeV, the alignment limit is always out of reach as long as the signal is $pp \to A\to Zh\to Zbb$ because it vanishes at $\cos(\beta-\alpha)=0$ independent of the Higgs boson mass.  
The case of heavy Higgs boson conversion through $A\to ZH$ analyzed in \cite{cms1,atlas1,atlas2} is suitable at the alignment limit as the coupling is $\sin(\beta-\alpha)$. Therefore, for a given $\tan\beta$, the two production processes, i.e., $A\to ZH$ and $A\to Zh$ are complementary along the $\cos(\beta-\alpha)$ axis. 

In \cite{cms1} results are presented only for type 2 for a specific choice of the Higgs boson masses, i.e., $m_H=379$ GeV, $m_A=172$ GeV. The alignment limit is not reachable by the analysis presented in \cite{atlas1} due to the choice of $H \to WW$ which vanishes at $\cos(\beta-\alpha)=0$. The analysis reported in \cite{atlas2} is performed at the alignment limit but is limited to $m_A-m_H\geq m_Z$. Therefore scenarios with degenerate Higgs boson masses ($m_H\simeq mA$) are out of reach in \cite{atlas2} due to the choice of the signal .
  
One of the reasons for using $A\to Zh$ by the LHC collaborations is less number of free parameters in the signal due to the fixed value of the $h_{\mathrm{SM}}$ mass. Moreover, $A\to Zh$ can be tested for $A$ masses as low as $m_h+m_Z$ while such masses are not allowed in $A\to ZH$ due to $m_H>m_h$ assumption. However, as long as the alignment limit and its nearby area is concerned, $A\to ZH$ provides a higher sensitivity near $\cos(\beta-\alpha)=0$.

There are also differences in the CP-even Higgs boson decays to fermions as well as gauge bosons. Since $h\to bb$ has been analyzed by LHC, we will discuss about $H\to bb$ in the next sections. Concerning the heavy Higgs boson decay to gauge bosons, it was mentioned that the coupling, normalized to the corresponding SM value, is $\cos(\beta-\alpha)$ (Eq. \ref{hgauge}). Therefore combining $A\to ZH$ and $H\to VV$ ($V ~=~ W~ \mathrm{or}~ Z$) may not be a reasonable idea as the two coupling factors compensate each other and the higher the production cross section, the lower the $H\to VV$ decay rate. This is not the case for $A\to Zh$ followed by $h\to VV$ as both involve $\cos(\beta-\alpha)$ factors. However, it is essentially a production chain most suitable far from the alignment area. On the other hand, the higher final state particle multiplicity due to gauge boson decays leads to no superiority over fermionic final states.

Therefore the conclusion for both processes ($A\to ZH$ and $A\to Zh$) is to preferably use $H/h\to f\bar{f}$. Here, we denote the decay final state as $f\bar{f}$ to remember that in lepton-specific type, the $\tau\tau$ final state has to be used while in other types $b\bar{b}$ is the most suitable final state of the light or heavy CP-even Higgs boson.

Another possibility, which is currently missing among the list of LHC analyses, is to use single neutral gauge boson production leading to the Higgs boson pair production, i.e., $pp~ \mathrm{or}~e^+e^- \to Z^* \to AH$. Here we also include lepton collisions at future colliders. The final state can be set by $A \to bb,~H\to bb$ or $A\to bb, ~H\to VV$ with $b$ replaced by $\tau$ for the lepton-specific type. The $A\to VV$ can not be considered due to vanishing CP-odd Higgs-gauge coupling. Figures \ref{AHffff} and \ref{AHWWff} show the Feynman diagrams related to the Higgs boson pair production in the two final states discussed above. These are example diagrams for lepton pair collision and $V~=~W$ while for the case of LHC, the same signal is initiated through quark anti-quark annihilation.

Since the signal is proposed to be analyzed in the four $b$- or $\tau$-jet final state, reasonable control of the QCD background at the LHC event environment is crucial. However, we have shown in a number of analyses that the signal of this process can well be observed at future lepton colliders (the most recent results are found in \cite{prd2021}).

In the following sections, taking the Higgs boson pair production as the golden channel for extra Higgs boson studies, we discuss about the branching ratio of CP-even and CP-odd Higgs boson decays and reach the final conclusion by analyzing all main combinations of decays.

\section{CP-even heavy Higgs boson decay}
The Higgs boson pair production should be analyzed in a specific final state. The decay of the CP-even Higgs boson can occur in fermionic mode if the Higgs boson mass is below the threshold of the lightest gauge boson pair production, i.e., if $m_H<2m_W$. With $m_H>2m_W$, decay to $WW$ and then $ZZ$ (if $m_H>2m_Z$) are kinematically allowed. However, one should be aware of possible Higgs boson conversion, i.e., $H\to hh$, which turns on if $m_H>2m_h$. Therefore we discuss about the three regions as follows.
\subsection{$m_H<2m_W$}
In this region the Higgs boson decays to fermions, i.e., $b$ quarks in types 1 to 3, and $\tau$ leptons in type 4 (lepton-specific) unless decay to gauge bosons is enhanced by migrating from the alignment limit.

\begin{figure}[h]
\hspace*{-0.1in}
 \includegraphics[height=0.4\textwidth,width=0.5\textwidth]{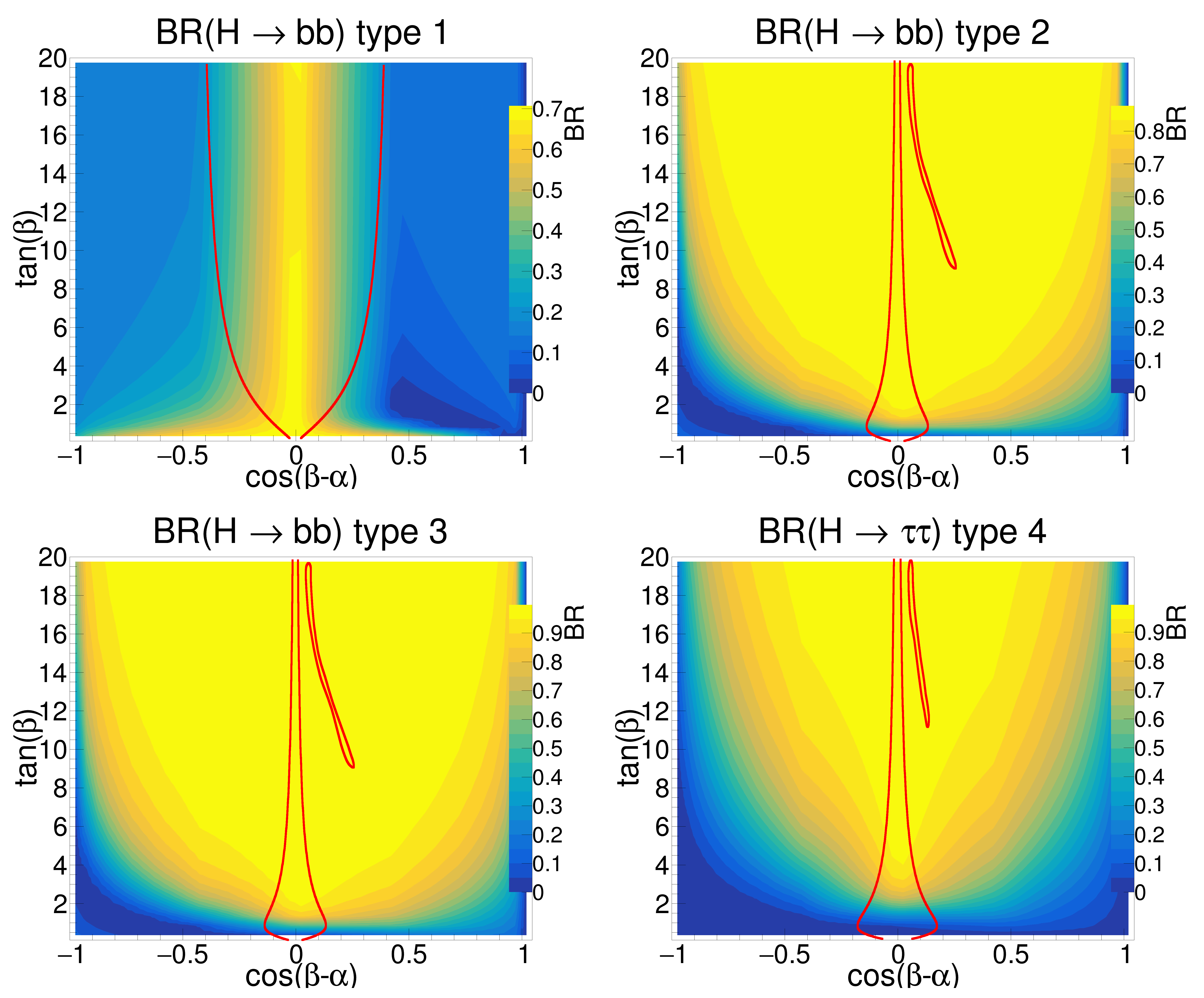}
 \caption{Branching ratio of Higgs boson decay to fermions ($bb$ final state in types 1 to 3 and $\tau\tau$ in type 4). The Higgs boson mass is set to 150 GeV. The area outside the red line is excluded by the LHC SM Higgs boson measurements. In type 2, the whole region is excluded by direct searches for heavy neutral Higgs boson. \label{Hff150}}
 \end{figure}

As seen in Fig. \ref{Hff150}, in type 1, $H\to bb$ is dominant near the alignment limit where $H\to VV$ is suppressed. However, increasing $|\cos(\beta-\alpha)|$ enhances $H\to VV$ in off-shell mode resulting in reduction of $H\to bb$ down to 0.2 or lower. However, the lower the Higgs boson mass, the higher the suppression of $H\to VV$. The two similar types 2 and 3 allow $H\to bb$ to be dominant in a wider area of the parameter space due to the $\tan\beta$ factor in the $H\to bb$ coupling (Eq. \ref{trig}). The type 4 behaves similar to type 1 with $b$ replaced by $\tau$. However, contrary to type 1 in which $H\to bb$ is almost $\tan\beta$ independent, in type 4, $H\to \tau\tau$ is enhanced at high $\tan\beta$.

The parameter space of the type 2 is very limited as it makes the Higgs sector of MSSM and is affected by those searches. The Higgs boson masses of 150 and 200 GeV are fully excluded at this type but at higher masses, the parameter space opens as verified by \texttt{HiggsBounds/HiggsSignals}.
  
\begin{figure}[h]
\hspace*{-0.1in}
 \includegraphics[height=0.4\textwidth,width=0.5\textwidth]{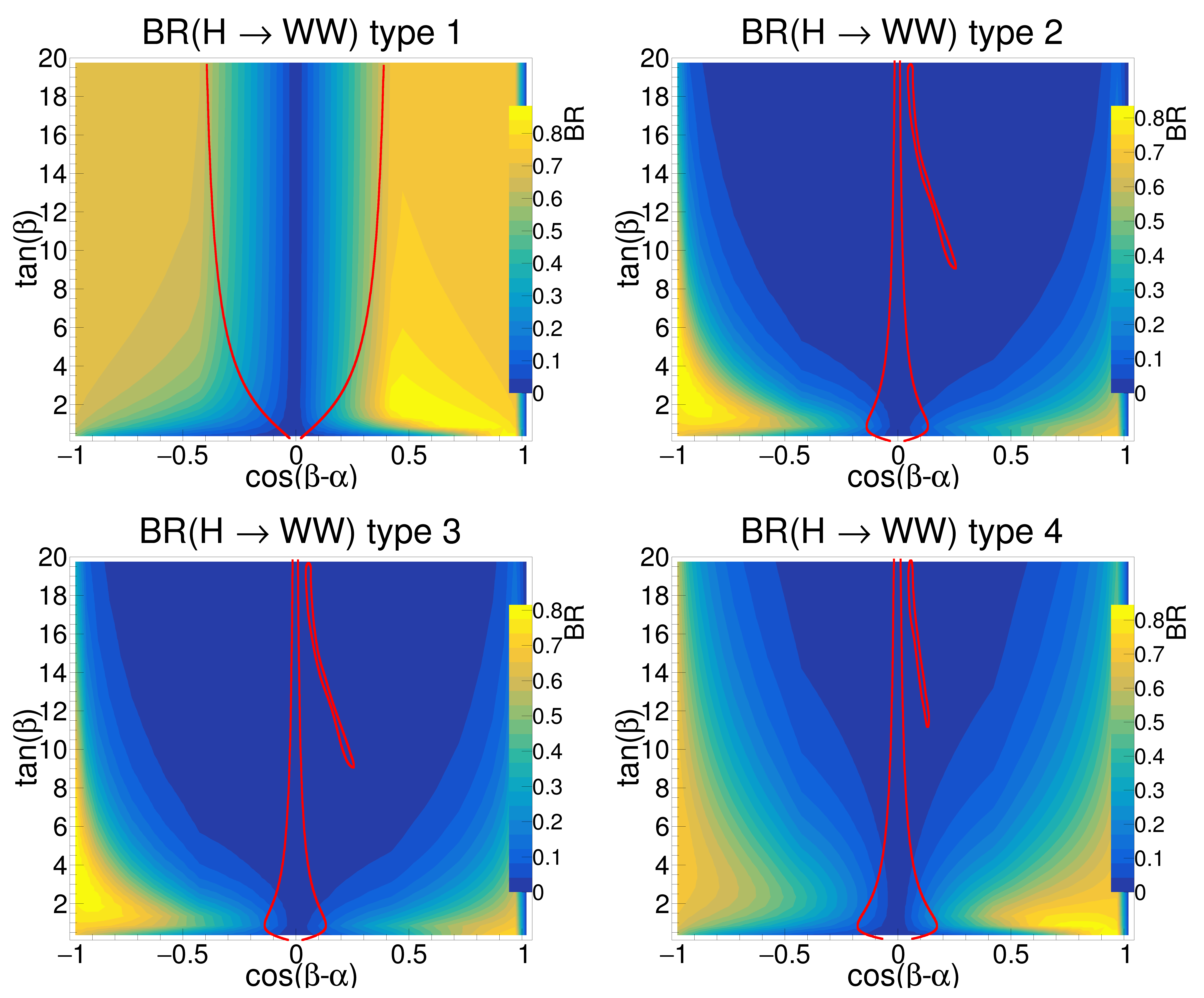}
\caption{Branching ratio of Higgs boson decay to $W$ boson pair assuming $m_H~=~150$ GeV. The area outside the red line is excluded. \label{HWW150}}
 \end{figure}

In order to compare the two fermionic and bosonic decay modes, we plot $H\to WW$ in the same parameter space as shown in Fig. \ref{HWW150}. The two complementary plots shown in Figs. \ref{Hff150} and \ref{HWW150} show how the two decay modes $H\to ff$ and $H\to WW$ compete. These two plots assume $m_H~=~150$ GeV.

Here we verify that the region of parameter space shown in Figs. \ref{Hff150} and \ref{HWW150} is theoretically accessible within the requirements of unitarity, stability and perturbativity.

\begin{figure}[h]
\hspace*{-0.1in}
 \includegraphics[height=0.4\textwidth,width=0.5\textwidth]{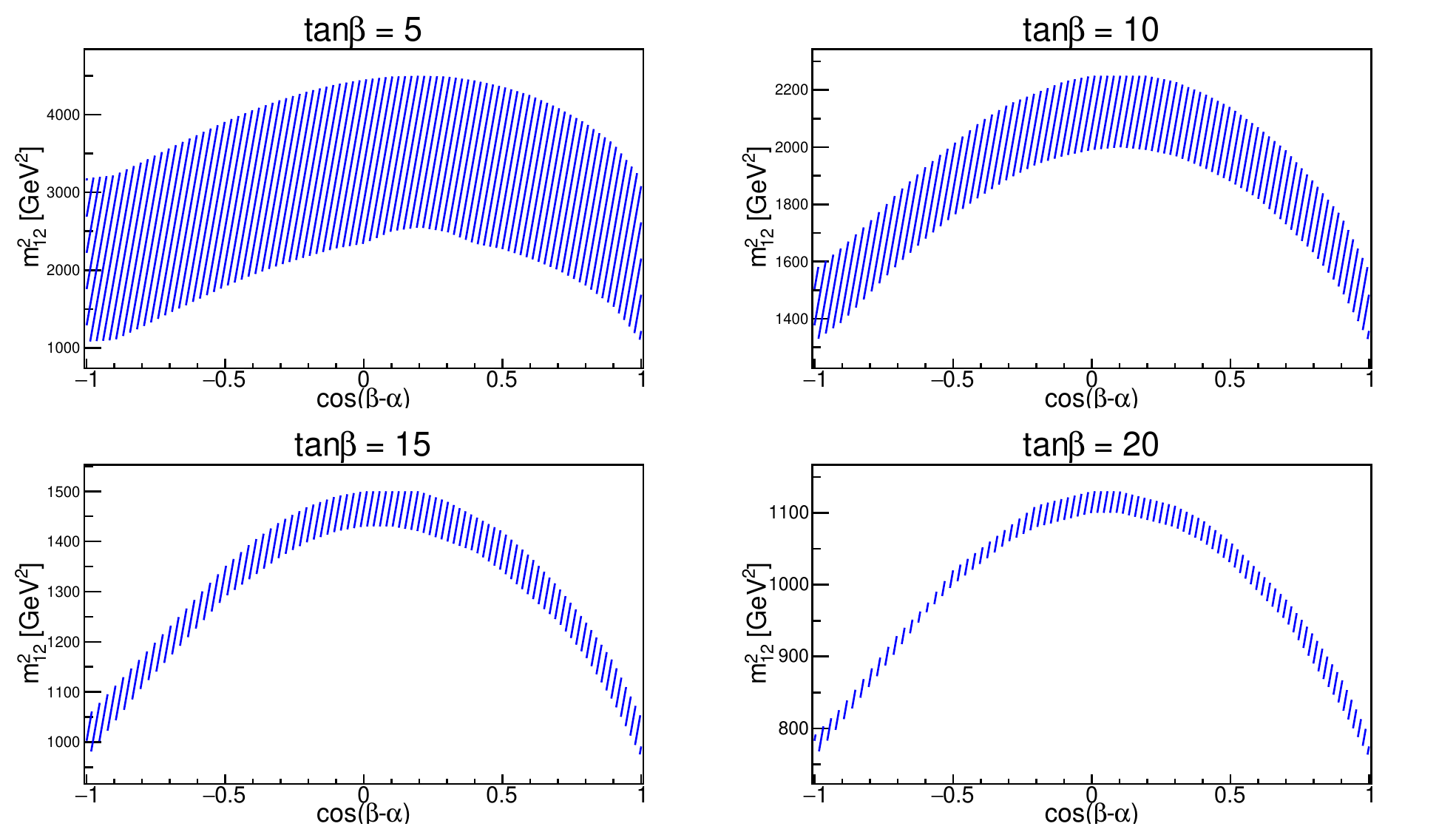}
\caption{The range of $m_{12}^2$ which satisfy theoretical constraints as a function of $\cos(\beta-\alpha)$ for the four values of $\tan\beta=5,~10,~15,~20$. The Higgs boson mass is set to $m_{H/A/H^{\pm}}~=~150$ GeV. These results are independent of the type of the model. \label{m12150}}
\end{figure}

In order to do so, for each point in the parameter space, we search for a range of $m_{12}^2$ values which satisfy theoretical constraints. Results are type independent and are shown for the four chosen values of $\tan\beta=5,~10,~15$ and 20 in Fig. \ref{m12150}. As is seen, increasing $\tan\beta$ shrinks the available $m_{12}^2$ range for a given point defined by the values of $\cos(\beta-\alpha)$ and $\tan\beta$.

For the mass scenario adopted in this section, BR$(H\to ff)$ and BR$(H\to VV)$ are independent of $m_{12}^2$ and any value of $m_{12}^2$ can be picked up from the range shown in Fig. \ref{m12150}. However, plots shown in Fig. \ref{m12150} confirm that there is at least one such $m_{12}^2$ value for each point in the parameter space in the range of $\tan\beta$ and $\cos(\beta-\alpha)$ under study.

We also verify that the provided allowed ranges for $m^2_{12}$ are consistent with $h \to \gamma\gamma$ measurements at the LHC. The BR($h \to \gamma\gamma$) slightly depends on $m^2_{12}$. For example, with $\cos(\beta-\alpha)=0$ and $\tan\beta=5$, the minimum and maximum theoretically allowed values for $m^2_{12}$ are 2338 and 4447 GeV$^2$ (shown in Fig. \ref{m12150}) for $m_{H/A}=150$ GeV. The corresponding values for BR($h \to \gamma\gamma$) are $2.41\times10^{-3}$ and $2.55\times10^{-3}$. These values are allowed by the LHC observations as verified by \texttt{HiggsBounds/HiggsSignals}.  
\begin{figure}[h]
\hspace*{-0.1in}
 \includegraphics[height=0.25\textwidth,width=0.35\textwidth]{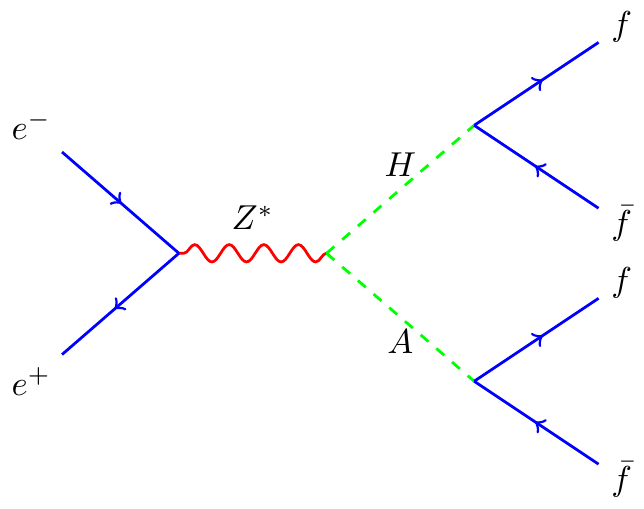}
 \caption{Higgs boson pair production in the four fermion final state. \label{AHffff}}
 \end{figure}
 \begin{figure}[h]
 \hspace*{-0.1in}
  \includegraphics[height=0.25\textwidth,width=0.35\textwidth]{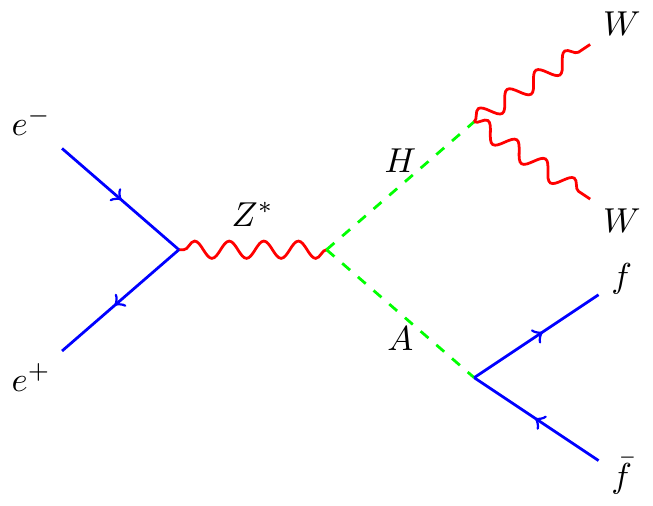}
  \caption{Higgs boson pair production in $WWff$ channel. Due to the large particle multiplicity in the final state, this channel may not provide better signal significance compared to the four fermion final state. \label{AHWWff}}
\end{figure}
\subsection{$2m_W<m_H<2m_h$}
In this region, $H\to WW$ starts to occur in on-shell mode and if $m_H>2m_Z$, $H\to ZZ$ will also be present. The relevant domain of $H\to VV$ is limited to Higgs boson masses below the threshold of SM-like Higgs boson pair production, i.e., $2m_W \lesssim m_H \lesssim 2m_h$.

\begin{figure}[h]
\hspace*{-0.1in}
 \includegraphics[height=0.4\textwidth,width=0.5\textwidth]{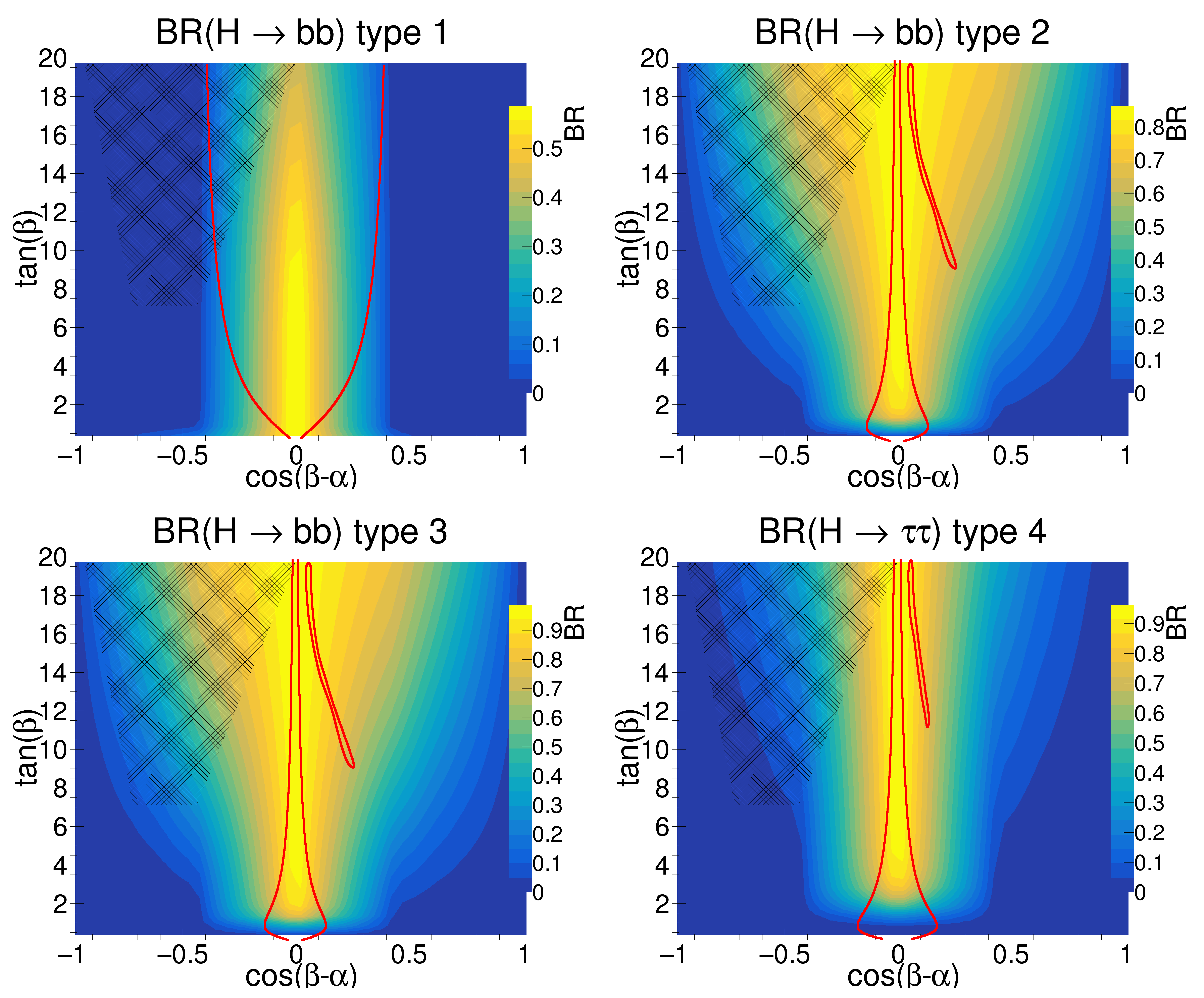}
\caption{Branching ratio of Higgs boson decay to fermion pair assuming $m_H~=~200$ GeV. The hashed region on the top left is theoretically inaccessible and the area outside the red line is excluded by the LHC SM Higgs boson measurements. In type 2, the whole region is excluded by direct searches for heavy neutral Higgs boson. \label{Hff200}}
\end{figure}

\begin{figure}[h]
\hspace*{-0.1in}
 \includegraphics[height=0.4\textwidth,width=0.5\textwidth]{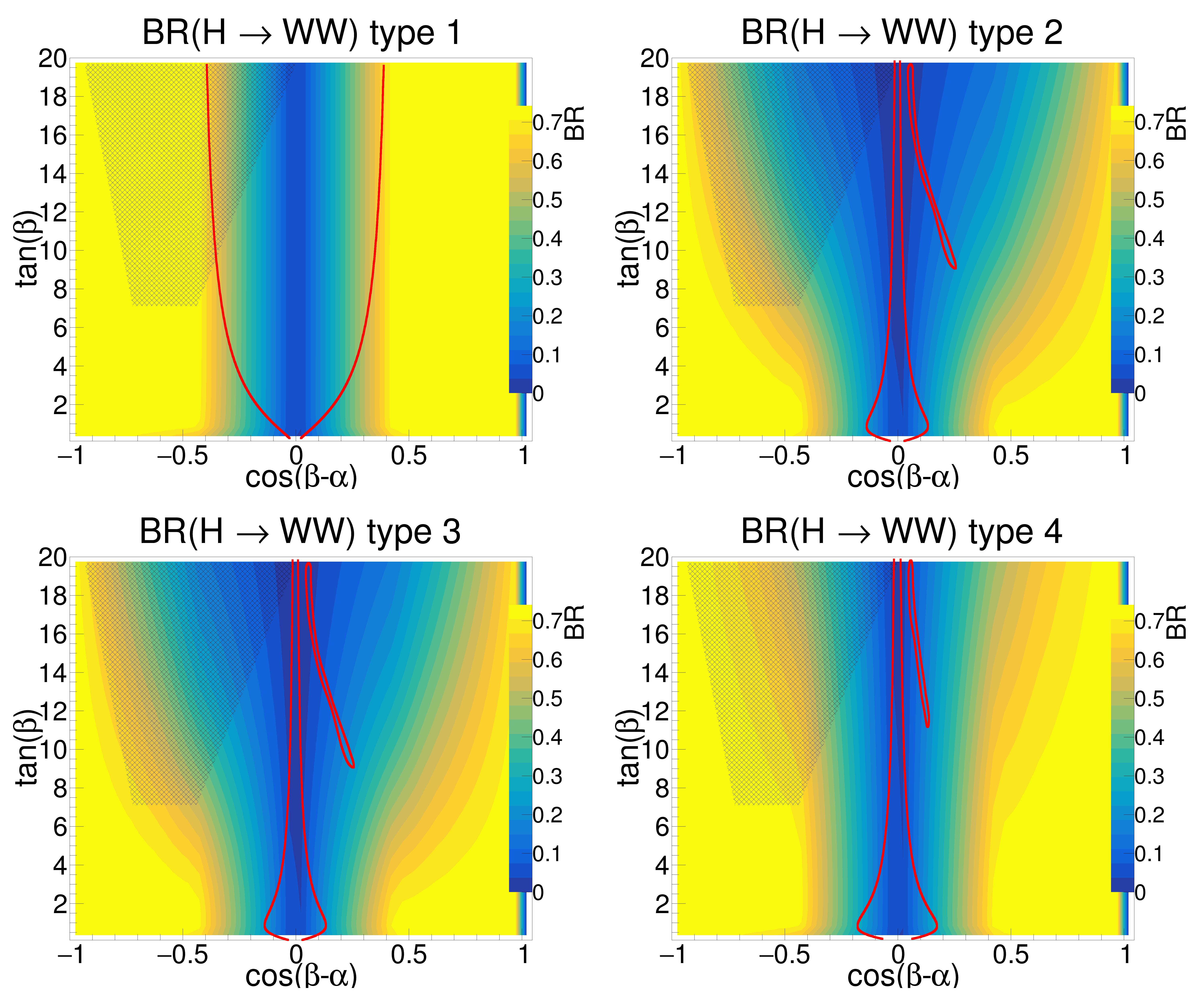}
\caption{Branching ratio of Higgs boson decay to $W$ boson pair assuming $m_H~=~200$ GeV. The hashed region on the top left is theoretically inaccessible and the area outside the red line is experimentally excluded. \label{HWW200}}
\end{figure}

As for illustration, we show BR($H\to ff$) and BR($H\to WW$) in Figs. \ref{Hff200} and \ref{HWW200} respectively. The Higgs boson mass is set to $m_H~=~200$ GeV. The Higgs boson decay to gauge boson pair is dominant when $|\cos(\beta-\alpha)|$ approaches unity unless $H\to ff$ is enhanced at high $\tan\beta$ values in types 2 to 4. Inverse colors in the two plots shown in Figs. \ref{Hff200} and \ref{HWW200} show that there is no other relevant decay mode for such Higgs boson masses in the range $2m_W \lesssim m_H \lesssim 2m_h$.

\begin{figure}[h]
\hspace*{-0.1in}
 \includegraphics[height=0.4\textwidth,width=0.5\textwidth]{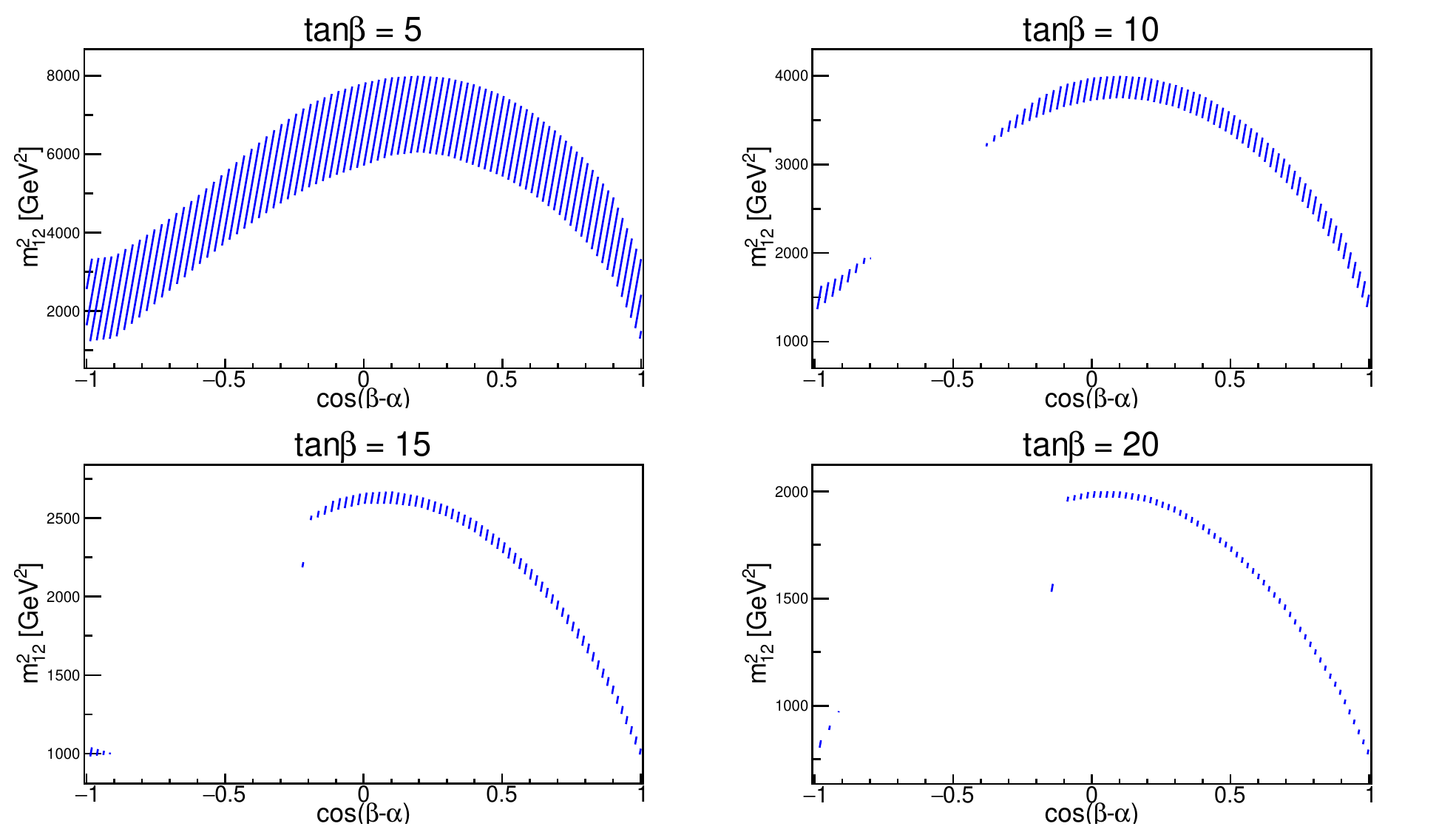}
\caption{The range of $m_{12}^2$ which satisfy theoretical constraints as a function of $\cos(\beta-\alpha)$ for the four values of $\tan\beta=5,~10,~15,~20$. The Higgs boson mass is set to $m_{H/A/H^{\pm}}~=~200$ GeV. These results are independent of the type of the model. \label{m12200}}
\end{figure}

The hashed region on the top left parts of Figs. \ref{Hff200} and \ref{HWW200} are excluded by theoretical constraints. The approach is the same as what was discussed in the  previous section. For a given $\tan\beta$ value, a range of $m_{12}^2$ is obtained under theoretical constraints. Fig. \ref{m12200} shows results for the four values of $\tan\beta$. In this case, at high $\tan\beta$, for some values of negative $\cos(\beta-\alpha)$ there is no $m_{12}^2$ value respecting theoretical constraints.
\subsection{$m_H>2m_h$}
If $m_H>2m_h$, i.e., with a Higgs boson mass above 250 GeV, there is possibility of $H\to hh$ with a type independent coupling which depends on $m_{12}^2$. The presence of this decay mode causes suppression of BR($H\to VV$). One needs to search for a range of $m_{12}^2$ which satisfy theoretical constraints. On the other hand, $m_{12}^2$ dependence of the $H\to hh$ coupling leads to dependence of branching ratio of all other decay modes especially $H\to VV$ on $m_{12}^2$ as the sum of all BRs has to be unity.

The only safe area in this mass region is the vertical line of $\cos(\beta-\alpha)=0$ and nearby where both $H\to VV$ and $H\to hh$ are suppressed and there is always a range of $m_{12}^2$ which respects theoretical constraints. Moreover, due to smallness of the above decay modes in the central region, BR$(H\to ff)$ is effectively independent of $m_{12}^2$.

\begin{figure}[h]
\hspace*{-0.1in}
 \includegraphics[height=0.4\textwidth,width=0.5\textwidth]{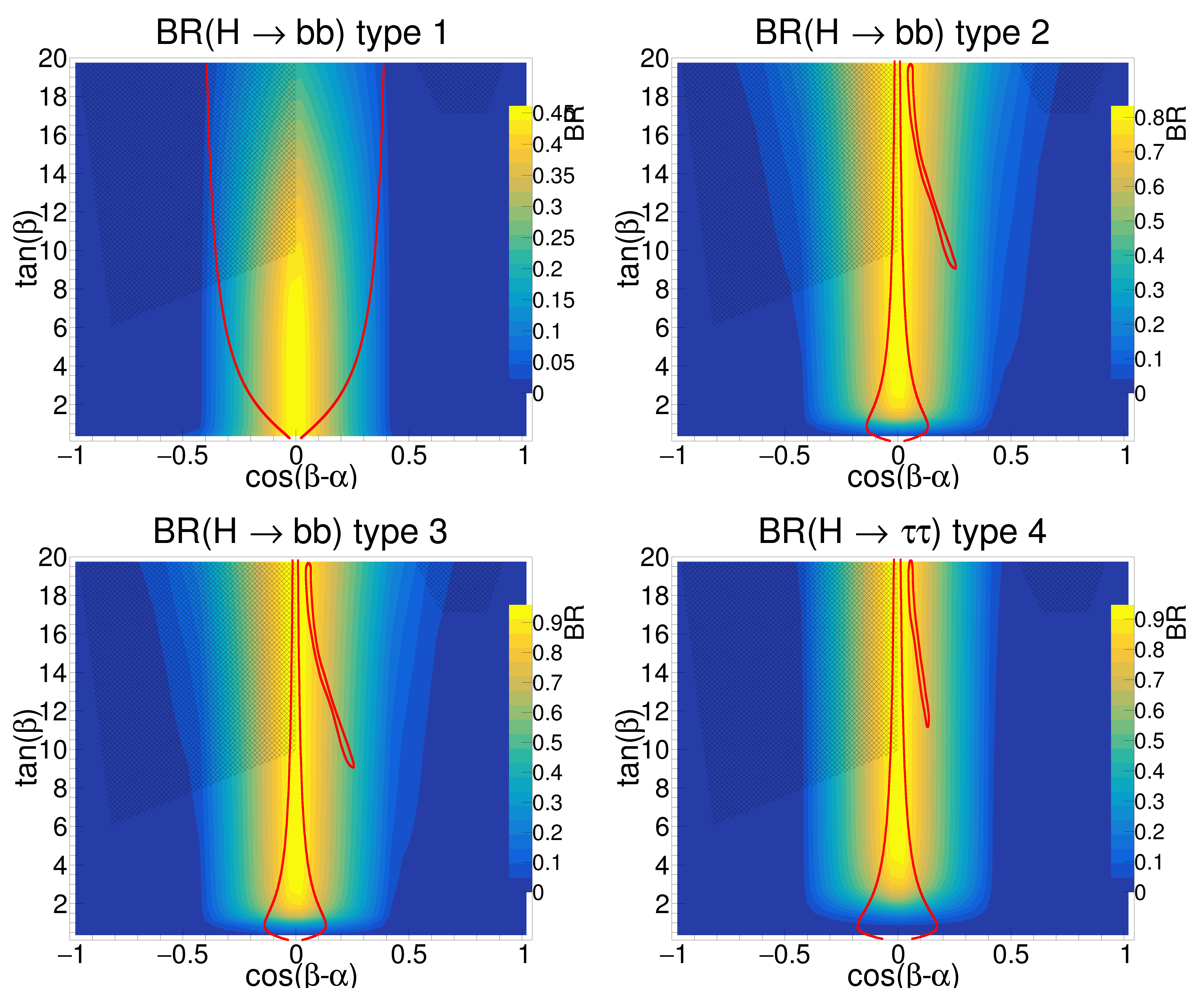}
\caption{Branching ratio of Higgs boson decay to fermion pair assuming $m_H~=~260$ GeV. The hashed regions on the top left and right are theoretically inaccessible and the area outside the red line is experimentally excluded. \label{Hff260}}
\end{figure}

Let us show BR($H\to ff$) for $m_H=260$ GeV in Fig. \ref{Hff260} which features dominance over other decay modes as well as $m_{12}^2$ independence at the alignment limit. The $m_{12}^2$ value has been set to 1000 GeV$^2$ in Fig. \ref{Hff260}. However, $m_{12}^2$ concerns rise when migrating from the alignment limit where BR($H\to ff$) is essentially small.

\begin{figure}[h]
\hspace*{-0.1in}
 \includegraphics[height=0.24\textwidth,width=0.35\textwidth]{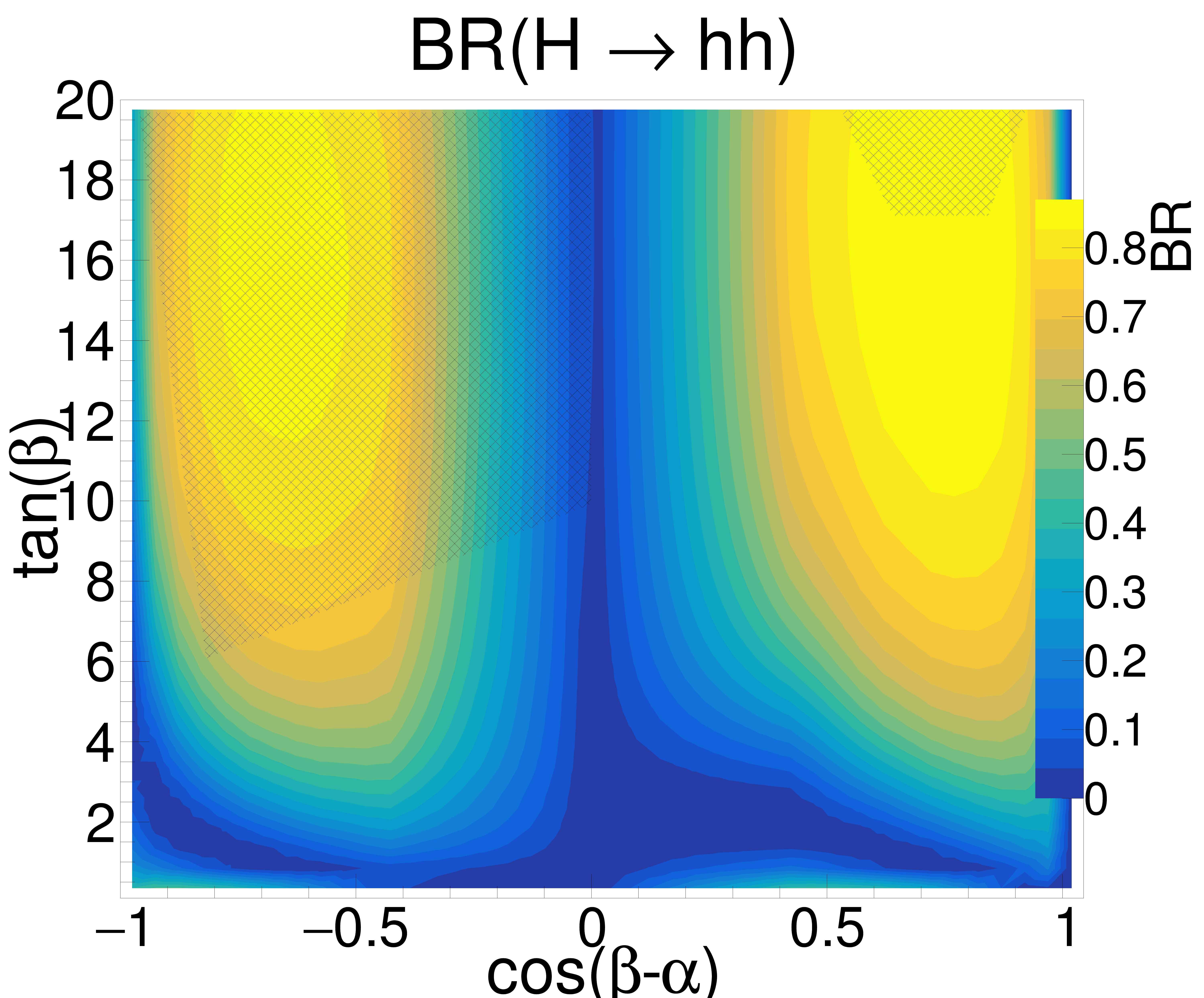}
\caption{Branching ratio of Higgs boson decay to SM-like Higgs boson pair assuming $m_H~=~260$ GeV. This is a type independent decay. However, the allowed regions are different for each type of the model and are shown by the red lines in Fig. \ref{Hff260}. \label{Hhh260}}
\end{figure}

The BR($H\to hh$) has been shown in Fig. \ref{Hhh260} again with fixed value of $m_{12}^2~=~$ 1000 GeV$^2$, which shows that $|\cos(\beta-\alpha)|>0$ area is under control of this decay mode except for the very low $\tan\beta$ values.

It is notable that there is a larger theoretically excluded area at this mass compared to the lower mass of 200 GeV. The excluded area at negative $\cos(\beta-\alpha)$ is larger and also extends to positive $\cos(\beta-\alpha)$ values at high $\tan\beta$. The dominant decay mode at $|\cos(\beta-\alpha)|>0.3$ is $H\to hh$  with BR$(H\to hh)>0.5$. At the central region of $\cos(\beta-\alpha)\simeq 0$, $H\to bb$ is still dominant as both $H\to hh$ and $H\to VV$ are suppressed when approaching this area.

There is a point in Fig. \ref{Hhh260} which should be cautious about. We plotted BR($H\to hh$) for a fixed value of $m_{12}^2$ to show the relevant domain of this decay mode in the parameter space. However, theoretical constraints rule out some parts of the plot in Fig. \ref{Hhh260} because for them the chosen $m_{12}^2$ may not be in the allowed range. Therefore for each point in the parameter space of $\tan\beta$ vs $\cos(\beta-\alpha)$ a value of $m_{12}^2$ should be picked up from the specified range to respect the theoretical constraints. The complexity is thus due to $m_{12}^2$ dependence of BR($H\to hh$) which provides a range of theoretically allowed branching ratios (not a single value) for each point in Fig. \ref{Hhh260}.

To conclude, study of $H\to VV$ at high masses faces difficulties due to theoretical considerations as well as the presence of other decay modes. However, the alignment limit can still be analyzed by $H\to ff$ at high masses.

For completeness, we note that if the Higgs boson mass is above the kinematic threshold of decay to top quark pair, $H/A \to t\bar{t}$ is also switched on. However, Higgs-top quark coupling in all types of the model is proportional to $\cot\beta$ and is suitable to be studied in type 1 where all Higgs-fermion couplings are proportional to $\cot\beta$ and cancel out in branching ratio of Higgs boson decays. In this case $H/A \to t\bar{t}$ is dominant when $m_H>2m_{top}$. In the same Higgs boson mass region, in types 2 and 3, $H/A \to b\bar{b}$ is dominant at high $\tan\beta$ values, while in type 4, $H/A \to \tau\tau$ will be the most promising decay mode.

\section{CP-odd heavy Higgs boson decay}
\begin{figure}[h]
\hspace*{-0.1in}
 \includegraphics[height=0.24\textwidth,width=0.35\textwidth]{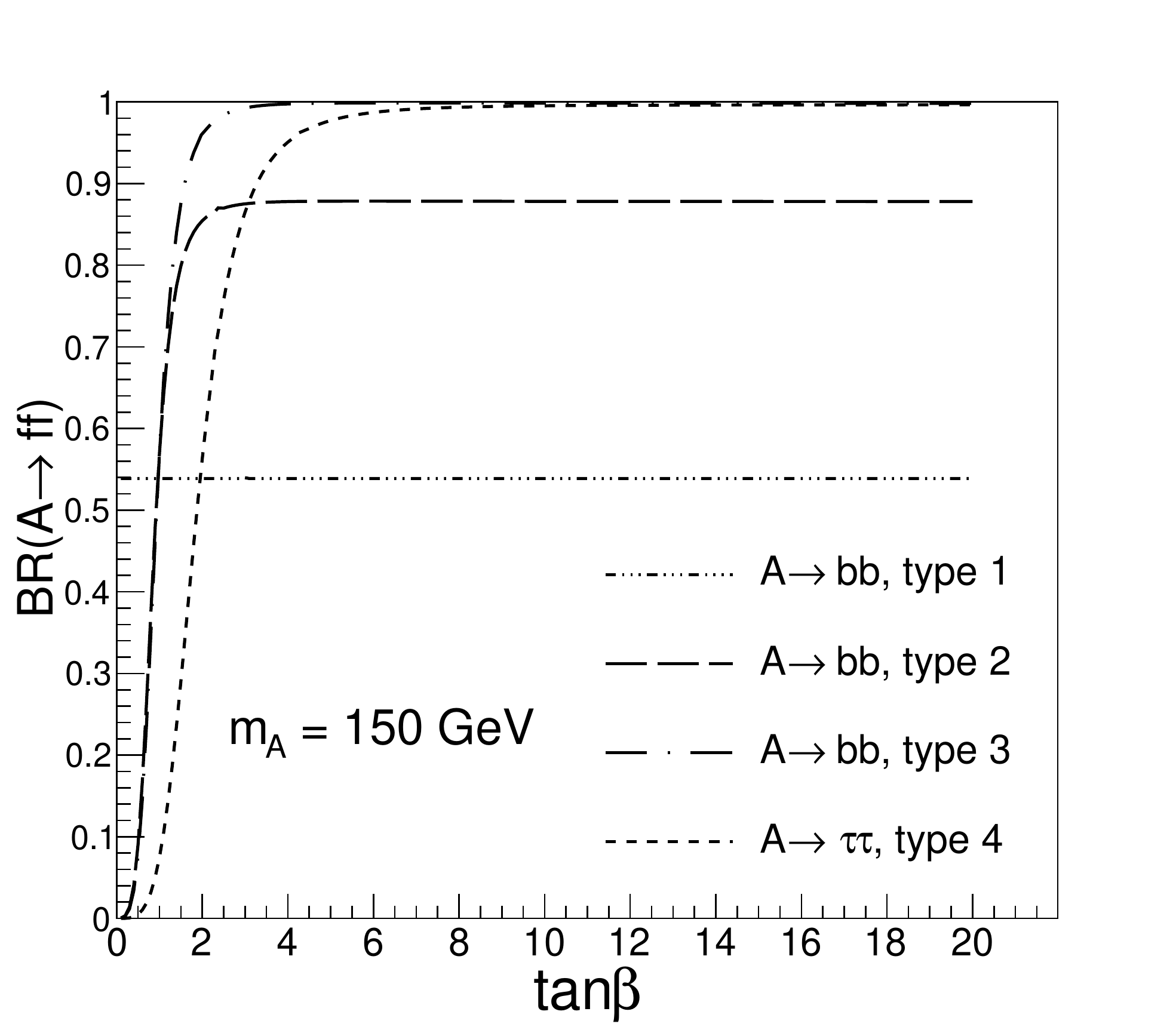}
\caption{Branching ratio of CP-odd Higgs boson decay to $bb$ in types 1 to 3 and $\tau\tau$ in type 4. The Higgs boson mass is set to 150 GeV. \label{Aff150}}
\end{figure}

\begin{figure}[h]
\hspace*{-0.1in}
 \includegraphics[height=0.24\textwidth,width=0.35\textwidth]{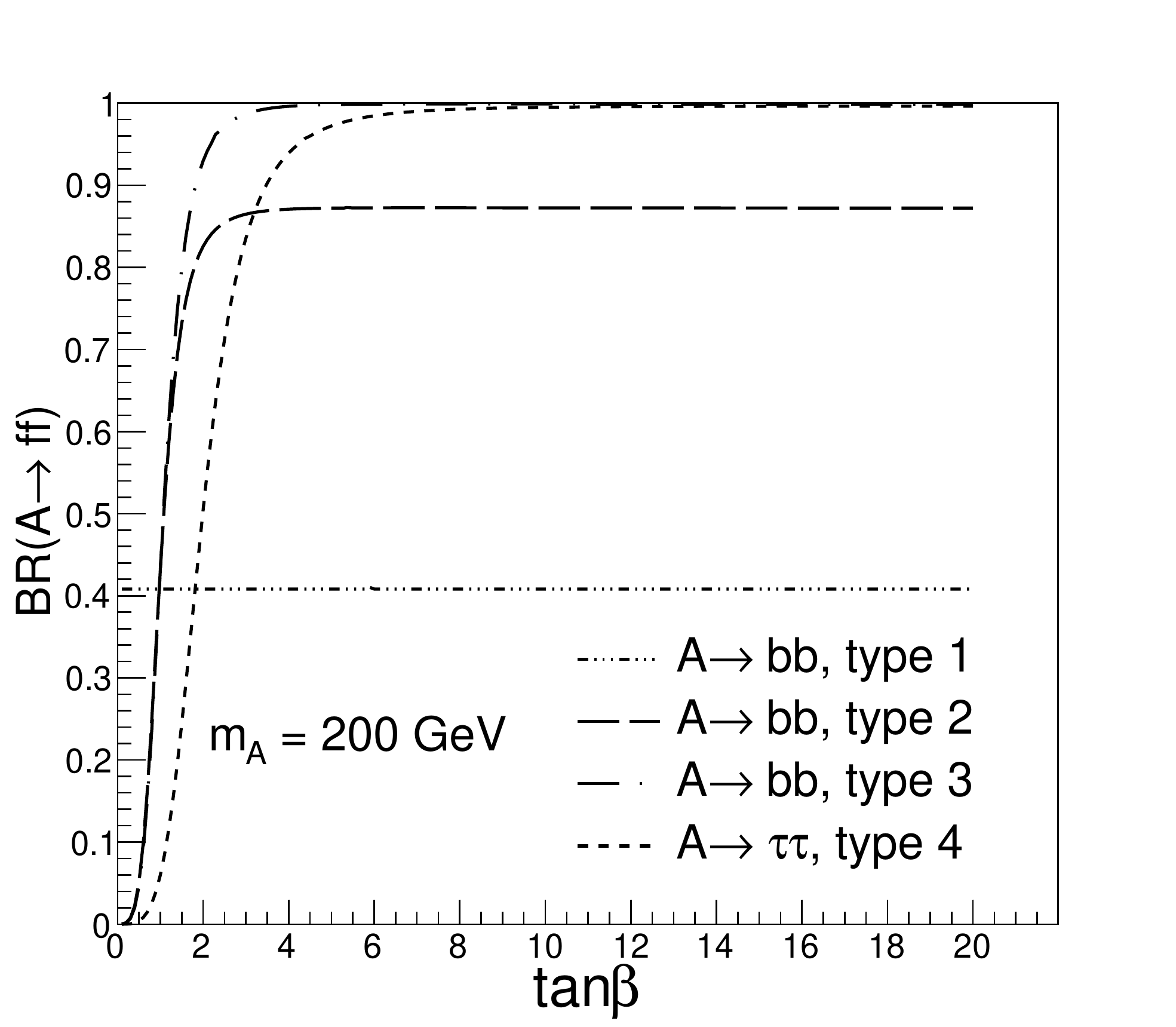}
\caption{Branching ratio of CP-odd Higgs boson decay to $bb$ in types 1 to 3 and $\tau\tau$ in type 4. The Higgs boson mass is set to 200 GeV. \label{Aff200}}
\end{figure}

The situation with CP-odd Higgs boson decays is simpler as $A\to VV$ vanishes and $A\to ff$ depends only on $\beta$ through $\tan\beta$ or $\cot\beta$. Therefore BR($A\to ff$) can be plotted as a function of $\tan\beta$ as shown in Figs. \ref{Aff150} and \ref{Aff200} for the two masses $m_A~=~150$ and 200 GeV. The main difference between the two masses is observed in type 1, where $A\to bb$ is more suppressed by $A\to gg$ at $m_A~=~200$ GeV. The other types essentially prefer $A\to ff$ at $\tan\beta>5$.
\section{Cross section of the Higgs boson pair production}
The only missing element for analyzing the Higgs boson pair production in the final states shown in Figs. \ref{AHffff} and \ref{AHWWff} is now the total cross section of $HA$ production which should then be multiplied by branching ratios of Higgs boson decays.

The cross section of these events depends on $\sin(\beta-\alpha)$ through the second vertex and for a fixed value of $\sin(\beta-\alpha)$ is independent of $\tan\beta$. The cross section of the first mass scenario ($m_H=m_A=150$ GeV) can be calculated for FCC-ee center of mass energy of 365 GeV \cite{FCCEnergy} as well as ILC \cite{ILCEnergy,ILCEnergy2} and CLIC stage 1 \cite{cliccdr} operating at center of mass energy of 500 GeV. The second mass scenario above the vector boson pair production threshold can be realized at the same operation scenario of ILC and CLIC.

As previously mentioned, the signal cross section prefers the central region of the alignment due to the $\sin(\beta-\alpha)$ factor thus preferring $H\to ff$ over $H\to VV$ because at the alignment limit, in all mass scenarios mentioned before, $H\to ff$ is dominant.

\begin{figure}[t]
\hspace*{-0.1in}
 \includegraphics[height=0.4\textwidth,width=0.5\textwidth]{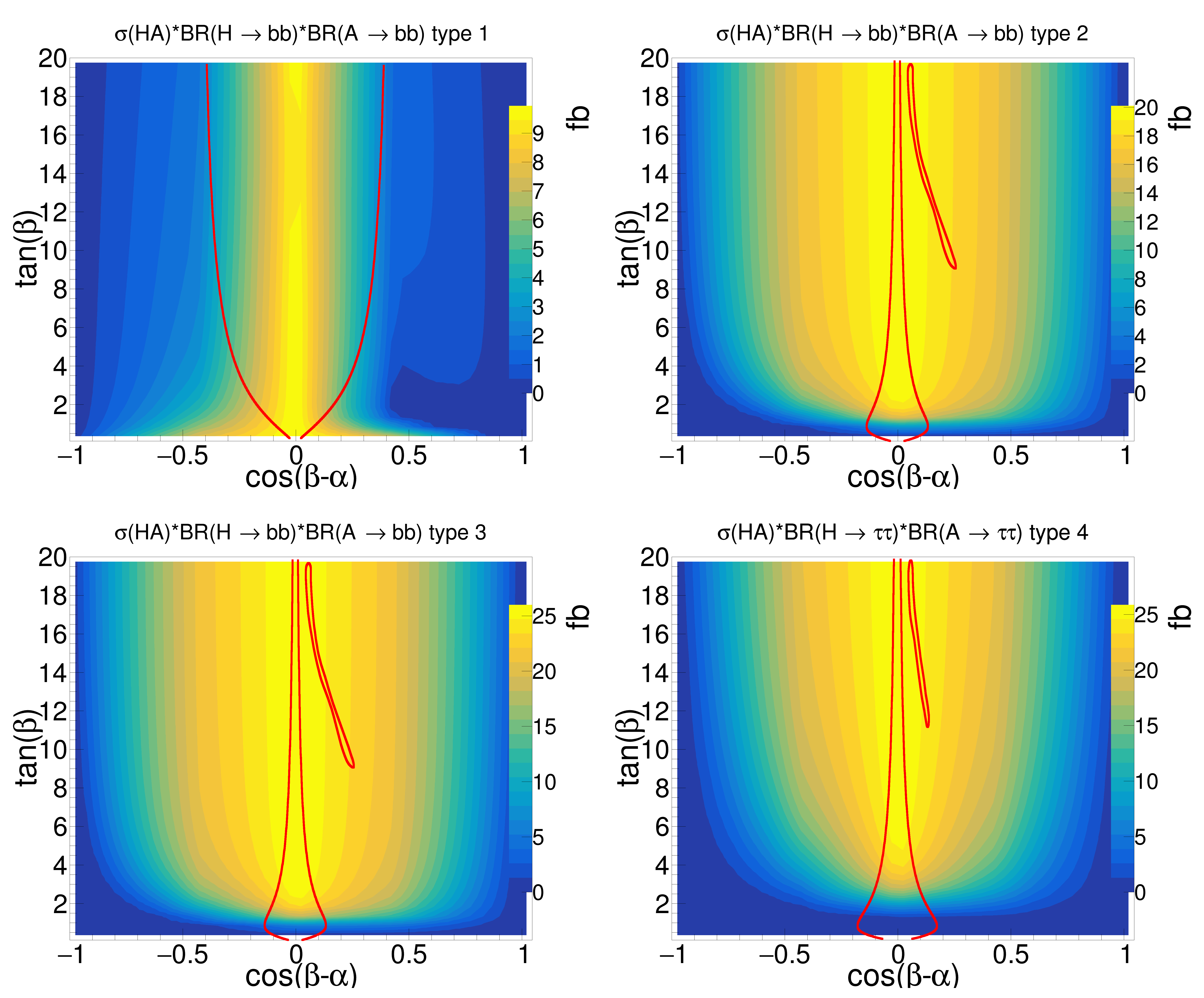}
\caption{The signal cross section in the four fermion final state for the mass scenario $m_H=m_A=150$ GeV at $\sqrt{s}=500$ GeV. \label{sigma1}}
\end{figure}

\begin{figure}[]
\hspace*{-0.1in}
 \includegraphics[height=0.4\textwidth,width=0.5\textwidth]{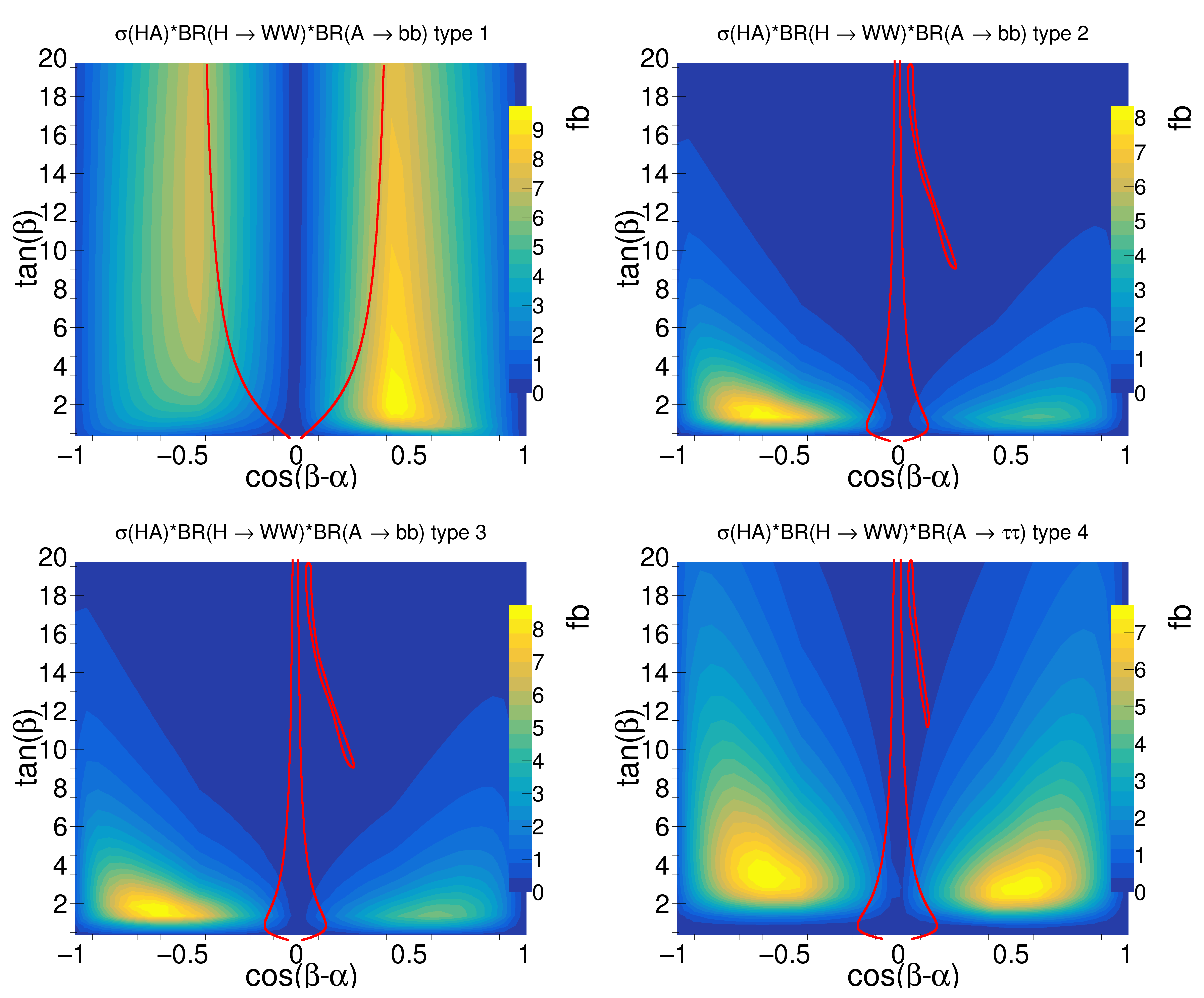}
\caption{The signal cross section in the $WWff$ final state for the mass scenario $m_H=m_A=150$ GeV at $\sqrt{s}=500$ GeV. \label{sigma2}}
\end{figure}

\begin{figure}[]
\hspace*{-0.1in}
 \includegraphics[height=0.4\textwidth,width=0.5\textwidth]{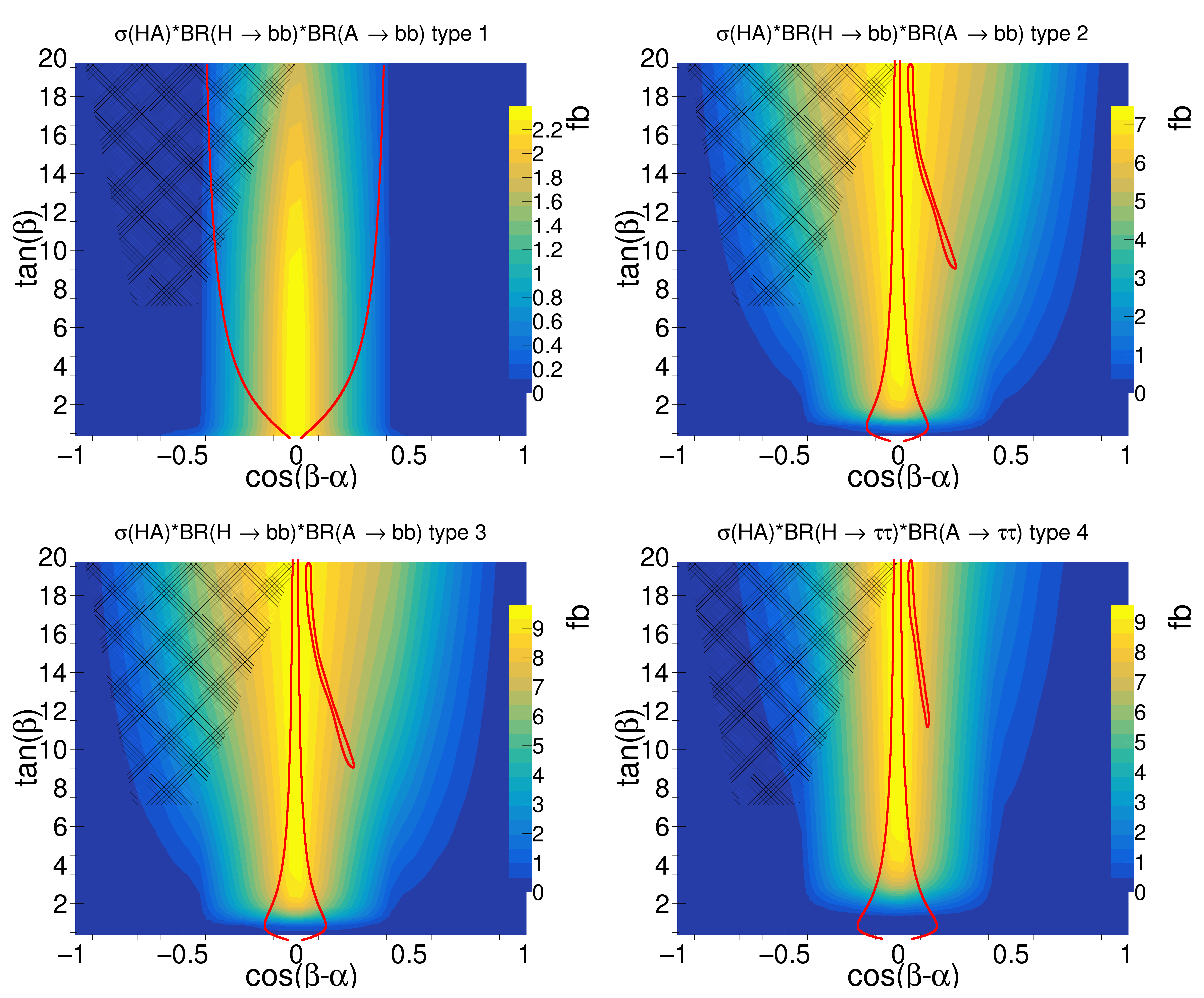}
\caption{The signal cross section in the four fermion final state for the mass scenario $m_H=m_A=200$ GeV at $\sqrt{s}=500$ GeV. \label{sigma3}}
\end{figure}

\begin{figure}[]
\hspace*{-0.1in}
 \includegraphics[height=0.4\textwidth,width=0.5\textwidth]{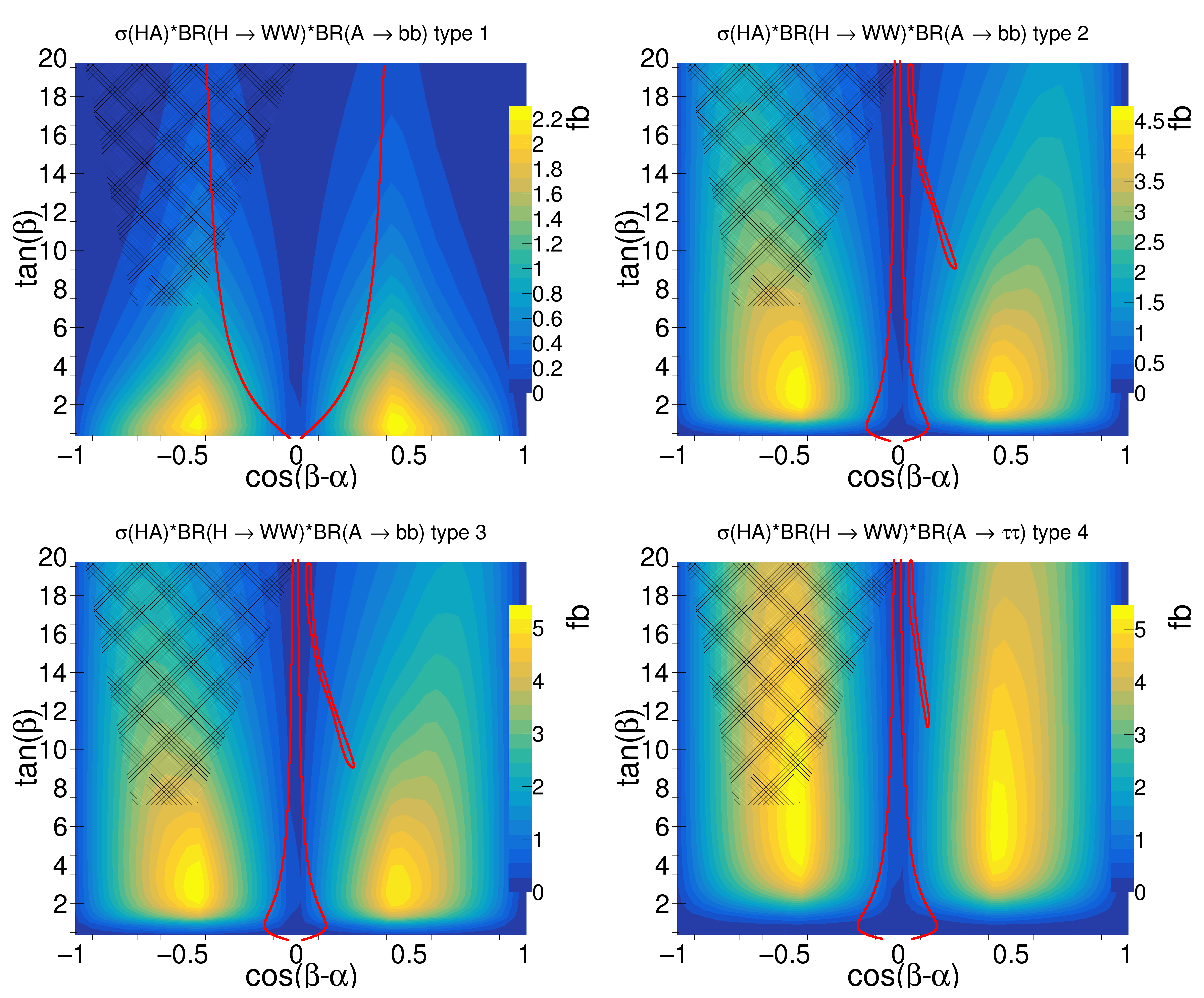}
\caption{The signal cross section in the $WWff$ final state for the mass scenario $m_H=m_A=200$ GeV at $\sqrt{s}=500$ GeV.\label{sigma4}}
\end{figure}

The product of cross sections and branching ratio of CP-even and CP-odd Higgs bosons are presented in Figs. \ref{sigma1} and \ref{sigma2} for the first mass scenario and Figs. \ref{sigma3} and \ref{sigma4} for the second mass scenario at CLIC/ILC center of mass energy $\sqrt{s}=500$ GeV for the two final states $H/A\to ff$ and $H\to WW,~A\to ff$. The color palette obviously shows that relevant final state for the central region is $H/A\to ff$ while regions far from the alignment limit can be probed by $H\to WW,~A\to ff$ with its own difficulties.
\section{Event Analysis at 500 GeV lepton collider}
We proceed to perform an event analysis at a lepton collider operating at $\sqrt{s}~=~500$ GeV. We do not check FCC center of mass energy due to the missing beam spectrum file for the $t\bar{t}$ operation scenario of $\sqrt{s}~=~365$ GeV.\\
The analysis is limited to the four jet final state where the jets are $b-$jets in types 1 to 3 and $\tau-$jets in type 4. The two scenarios of $m_{H}~=~m_{A}~=~150$ GeV and $m_{H}~=~m_{A}~=~200$ GeV are considered with event selection algorithm similar to what we presented in a previous work \cite{prd2021}.

The analysis is performed based on a single point in the parameter space of 2HDM defined by $\sin(\beta-\alpha)=1$ and $\tan\beta=10$. These parameters are used for the signal cross section and decay rates calculations but since event kinematics is independent of them, results can be easily scaled to other points in the parameter space.

The event generation is performed by \texttt{WHIZARD 3.0.0} \cite{whizard1,whizard2} and the beam spectrum of 500 GeV ILC is used to account for the ISR and beamstrahlung. The FSR and multi-particle showering is performed by \texttt{PYTHIA 8.3.03} \cite{pythia} followed by the detector simulation by \texttt{DELPHES 3.4.2} \cite{DELPHES} using \texttt{ILCGen} detector card. 

The hadronic background from photon interactions is taken into account by adding jet momentum smearing set to $0.3\%$ and $1.5\%$ for $|\eta|<0.76$ and $|\eta| \ge 0.76$ respectively based on the approach proposed by the CLIC collaboration \cite{overlay}.

The pseudorapidity is defined as $\eta=-\ln{\tan({\theta/2})}$ where $\theta$ is the polar angle with respect to the beam axis. 

The jet reconstruction is performed by \texttt{FASTJET 3.3.4} \cite{fastjet1,fastjet2} with anti-$k_{t}$ algorithm \cite{antikt} and the jet cone size $\Delta{R}=\sqrt{(\Delta{\eta})^{2}+(\Delta{\phi})^{2}}=0.5$. 

The jet tagging algorithms in \texttt{DELPHES} are based on MC truth matching with efficiencies depending on the jet energy and pseudorapidity for both $\tau-$ and $b-$tagging scenarios. As for the $b-$tagging we use average efficiency of 50$\%$ which was shown to work better against $t\bar{t}$ background in \cite{prd2021}. 

The event selection starts from choosing events with exactly four $b/\tau$-jets with $p_{T}>10$ GeV and $|\eta|<2$. We also perform a kinematic correction of the jet four momenta by correcting every jet four momentum component so that the four linear equations of momentum and energy conservation are satisfied. In order to have solutions for the set of four equations containing correction factors as unknowns, we consider a single correction factor for every jet and apply it on its four momentum components. Therefore the correction does not change the jet direction and only scales the four momentum vector.

\begin{figure}[]
	\hspace*{-0.1in}
	\includegraphics[height=0.4\textwidth,width=0.5\textwidth]{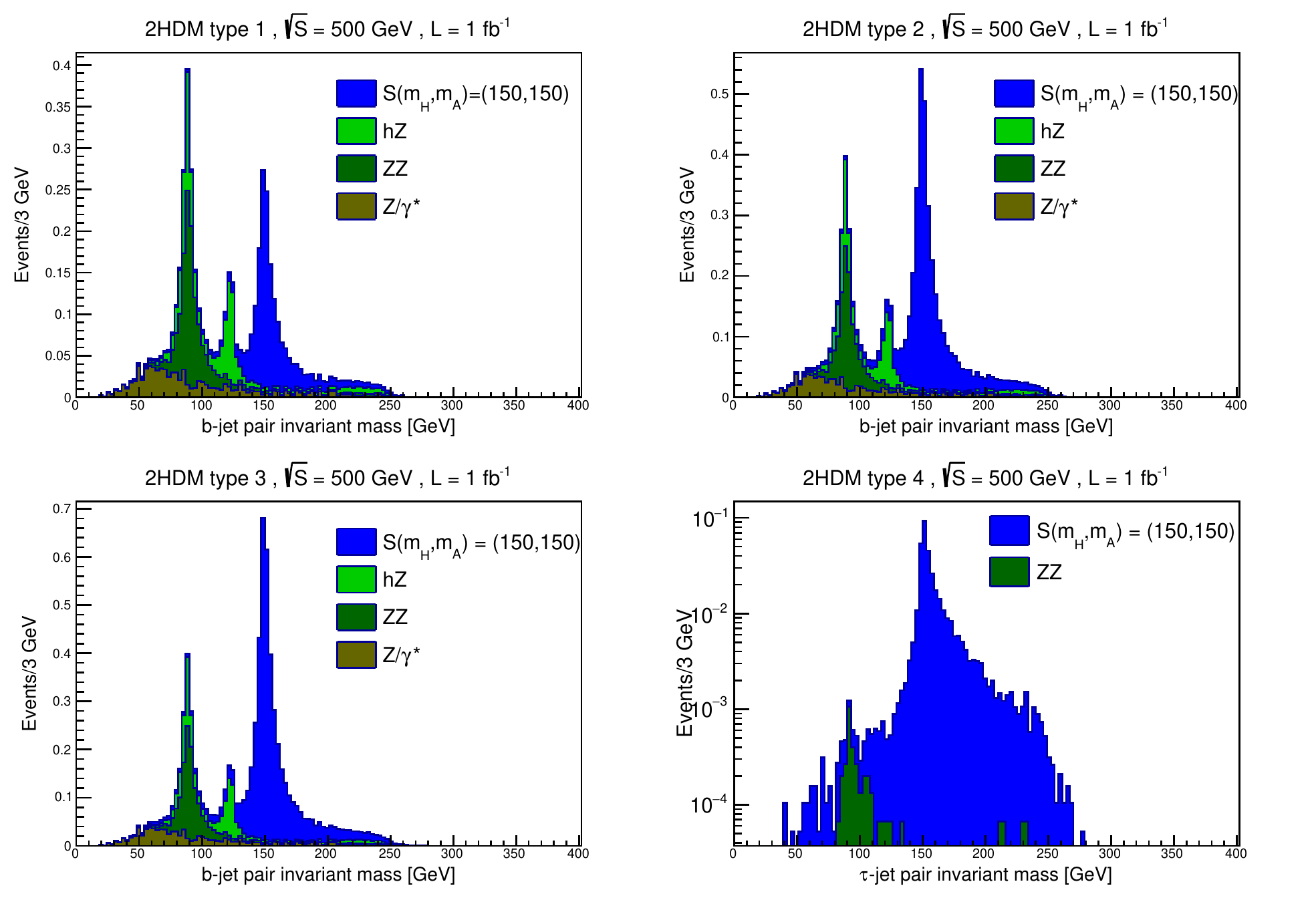}
	\caption{The invariant mass of the four jet final state in signal ($m_H=m_A=150$ GeV) and background events at $\sqrt{s}=500$ GeV normalized to 1 $fb^{-1}$. The model parameters are $\tan\beta=10$ and $\cos(\beta-\alpha)=0$.\label{BP1_4types}}
\end{figure}

\begin{figure}[]
	\hspace*{-0.1in}
	\includegraphics[height=0.4\textwidth,width=0.5\textwidth]{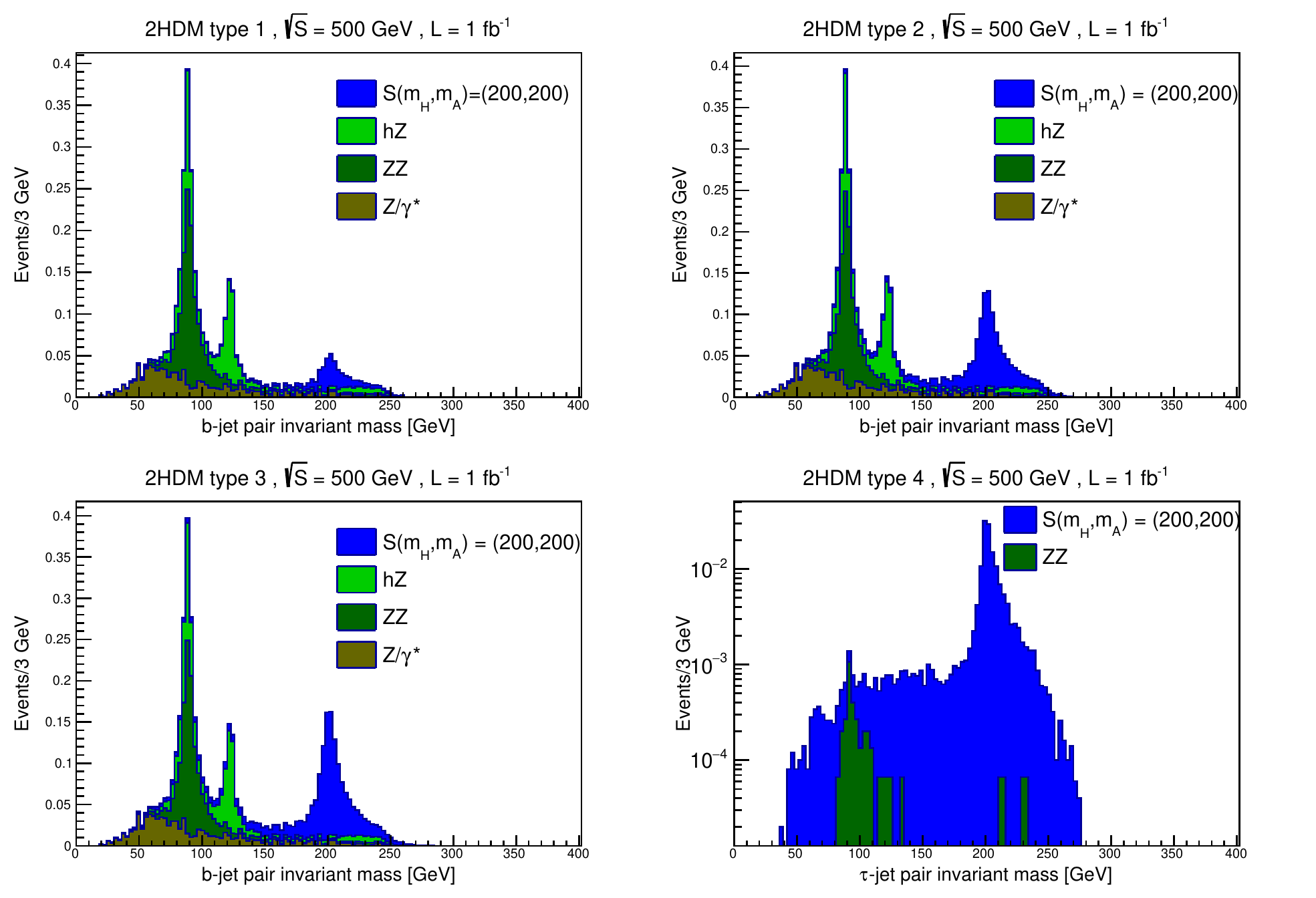}
	\caption{The invariant mass of the four jet final state in signal ($m_H=m_A=200$ GeV) and background events at $\sqrt{s}=500$ GeV normalized to 1 $fb^{-1}$. The model parameters are $\tan\beta=10$ and $\cos(\beta-\alpha)=0$.\label{BP2_4types}}
\end{figure}

After kinematic correction, jets are sorted in terms of their energies and the invariant mass of the second and third jets are calculated to make the signal distributions on top of the corresponding invariant mass distribution from the background events as shown in Figs. \ref{BP1_4types} and \ref{BP2_4types}. 

The idea of choosing the second and third jet is as follows. The two Higgs particles in the signal event produce their decay products back-to-back in their rest frames. However, each pair of jets fly at a specific angle which differs event to event. The two jets with the smaller angle with respect to the Higgs boson trajectory appear as the jets with maximum and minimum energies at the laboratory frame due to the Lorentz boost they receive. The other two jets from the other Higgs boson take the second and the third position in the energy-sorted list of jets. Here, results are shown based on using the second and the third jets, although similar results are obtained using the first and the fourth jets.       

Figures \ref{BP1_4types} and \ref{BP2_4types} show the distributions of the jet pair invariant masses in the two mass scenarios for the four types of 2HDM normalized to integrated luminosity of $1~fb^{-1}$. Only relevant background distributions are shown for each type. Other backgrounds like $WW$ and $t\bar{t}$ are negligible when the $b/\tau$-tagging is applied.

The $hZ$ background has also been shown assuming $m_h=125$ GeV. This is a variable background in the parameter space because BR($h \to bb/\tau\tau$) depends on $\tan\beta$ and $\cos(\beta-\alpha)$. The contribution of this background in the heavy Higgs boson mass window is very small.

In type 4, the SM background (which is mainly $ZZ$) is highly suppressed due to the low branching ratio of the $Z$ boson decay to $\tau\tau$ which is $\sim3\%$. This value has to be squared as there are two $Z$ bosons in such events. On the contrary, in signal events BR$(H/A \to \tau\tau)\sim1$.

 In other types of the model, we deal with $b$-jets final state with BR$(Z \to b\bar{b})\sim15\%$. There are also kinematic differences between the signal and $ZZ$ background related to the pseudo-rapidity distributions of the final state particles which were discussed in \cite{prd2021}. The lower efficiencies of $\tau$-tagging and related fake rate compared to the corresponding values for $b$-tagging and its fake rate also result in more suppression of the $\tau$-jets final state.  

The signal significance is obtained using the formula suitable for the low background statistics, i.e., $\sqrt{2(S+B)log(1+S/B)-2S}$ where $S$ and $B$ are the number of signal and background events at a given integrated luminosity inside the mass window. The mass window position and width is determined by maximizing the signal significance.

\begin{figure}[]
	\hspace*{-0.1in}
	\includegraphics[height=0.4\textwidth,width=0.5\textwidth]{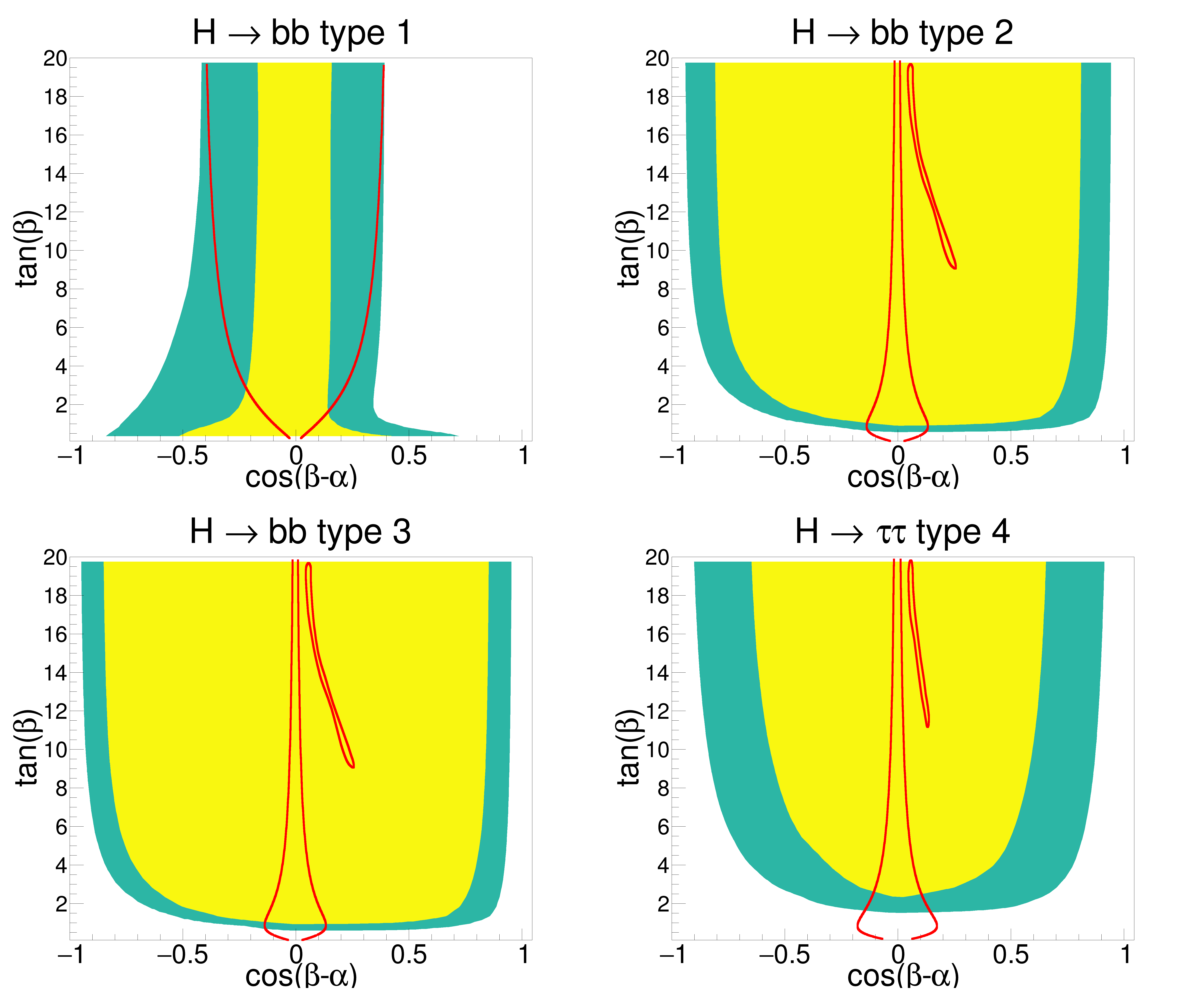}
	\caption{The 2$\sigma$ (95$\%$ C.L.) and 5$\sigma$ contours in green and yellow colors respectively for the mass scenario $m_H=m_A=150$ GeV ($\mathcal{L}= 10~ fb^{-1}$ and $\sqrt{s}=500$ GeV).\label{signif1}}
\end{figure}

\begin{figure}[]
	\hspace*{-0.1in}
	\includegraphics[height=0.4\textwidth,width=0.5\textwidth]{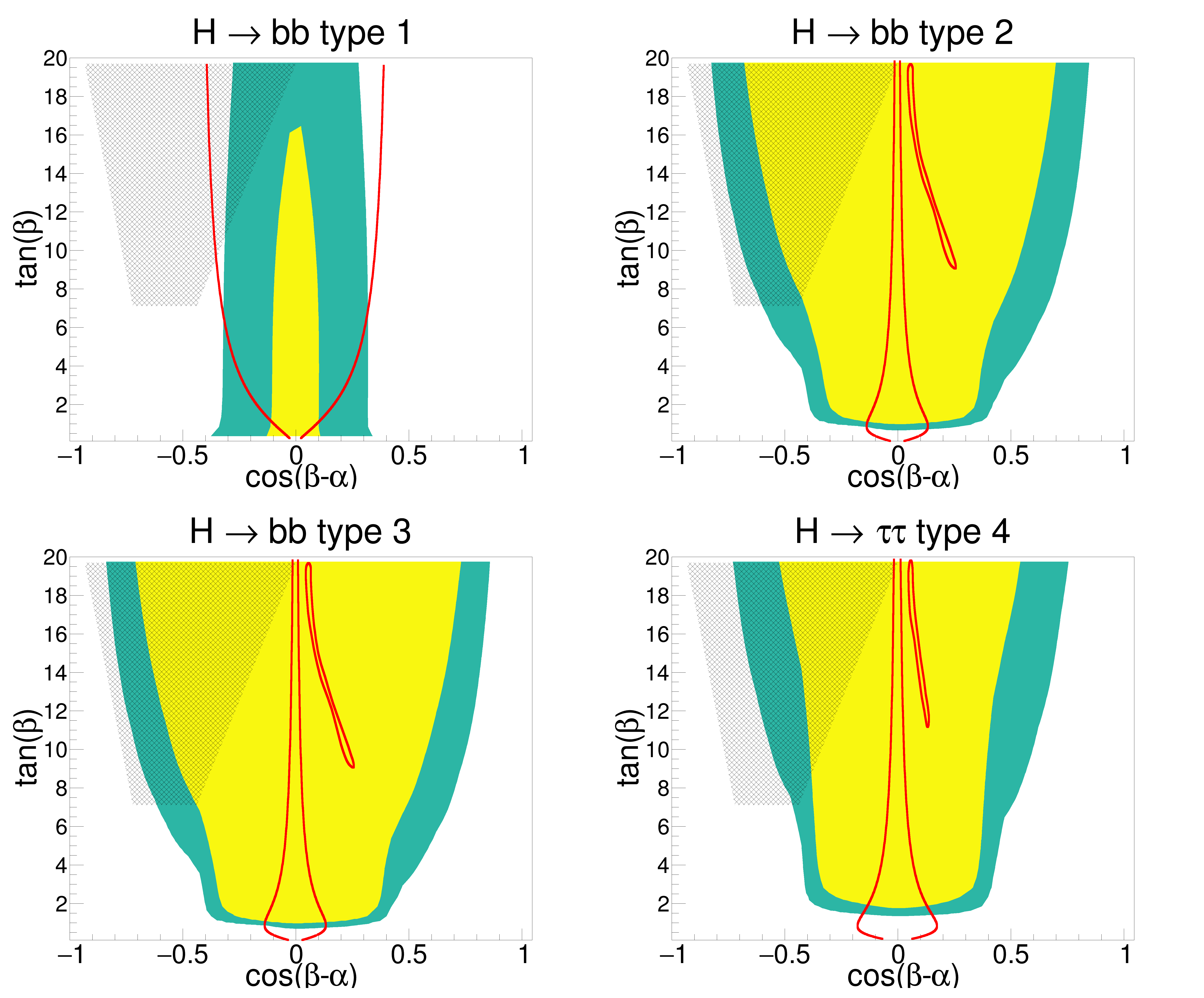}
	\caption{The 2$\sigma$ (95$\%$ C.L.) and 5$\sigma$ contours in green and yellow colors respectively for the mass scenario $m_H=m_A=200$ GeV ($\mathcal{L}= 100~ fb^{-1}$ and $\sqrt{s}=500$ GeV).\label{signif2}}
\end{figure}

Results for the two Higgs boson mass scenarios are shown in Figures \ref{signif1} and \ref{signif2}. The two regions shown in yellow and green are 5$\sigma$ and 2$\sigma$ contours respectively and the integrated luminosities of 10 $fb^{-1}$ and 100 $fb^{-1}$ have been assumed for the first and second mass scenarios. These amounts of data correspond to one or few weeks of operation of the collider. 
\pagebreak
\section{Conclusions}
The two Higgs doublet model was adopted as the theoretical framework for study of extra neutral Higgs bosons. The analysis was performed for all four types of the CP-conserving 2HDM and parameter space scans were presented including relevant parameters which determine the production cross sections and branching ratio of Higgs boson decays.

The results were divided into two domains of the Higgs boson mass, i.e., below and above the threshold of decay to gauge boson pair. Including experimental limits from the latest LHC results, it was shown that the $\cos(\beta-\alpha)=0$ limit known as the alignment limit is not yet excluded by LHC and can well be verified at lepton colliders if $e^+e^- \to HA\to 4$ fermion final state is analyzed. The reason is due to the dominance of the cross section as well as BR($H/A\to ff$) in this region. 

We also included an event analysis for the two mass scenarios and obtained the invariant mass distributions of signal and background events. Final results were presented as contours of 95$\%$ CL exclusion and 5$\sigma$ discovery for the four types. It is concluded that unexplored regions of 2HDM can well be excluded at 95$\%$ CL at $\mathcal{L}=$ 10 and 100 $fb^{-1}$ for the two mass scenarios respectively, which correspond to a week of collider operation or so. 
\begin{figure}[]
\hspace*{-0.1in}
 \includegraphics[height=0.24\textwidth,width=0.35\textwidth]{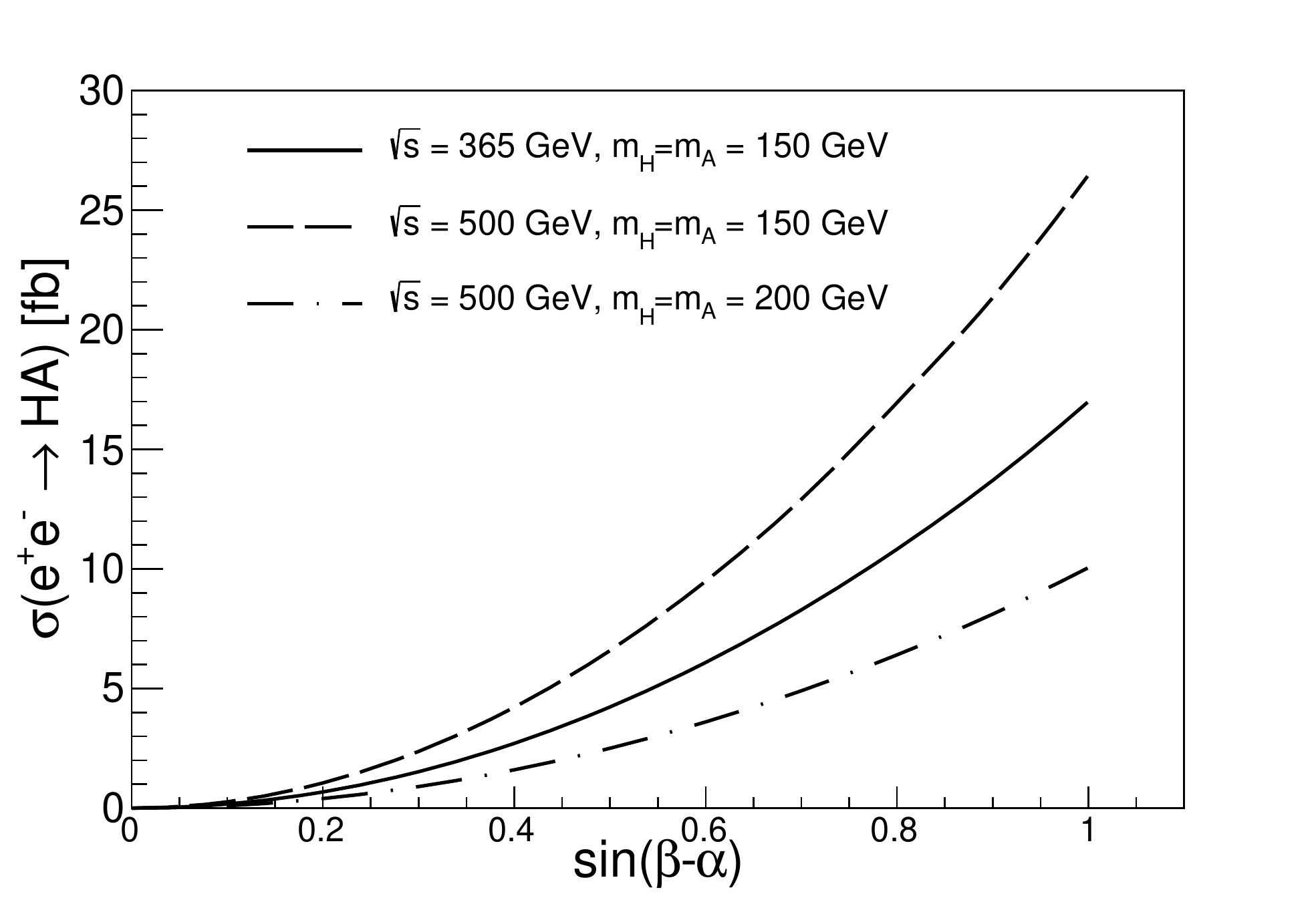}
\caption{The signal cross section as a function of $\sin(\beta-\alpha)$ for the two center of mass energies $\sqrt{s}~=~365$ GeV (FCC-ee) and 500 GeV (ILC or CLIC). The two scenarios of $m_{H}=m_{A}=150$ GeV and 200 GeV are shown. The signal cross section has a quadratic dependence on $\sin(\beta-\alpha)$ and reaches its maximum at $\sin(\beta-\alpha)=1$ for each mass scenario. \label{sigma}}
\end{figure}

%\bibliography{BIB_TO_USE}
%\bibliographystyle{apsrev}

\end{document}